%% file: main.tex
\documentclass[aps,prl,twocolumn,floatfix,superscriptaddress,nofootinbib,10pt]{revtex4-2}

\usepackage{amsmath,amssymb,amsfonts}
\usepackage[bb=boondox]{mathalfa}
\usepackage{graphicx}
\usepackage{bm}
\usepackage{amsmath}
\usepackage{amssymb}
\usepackage{amsfonts}
\usepackage{dsfont}
\usepackage{mathtools}
\usepackage{physics}
\usepackage{xcolor}
\usepackage{enumitem}
\usepackage[colorlinks=true, linkcolor=blue, citecolor=blue, urlcolor=blue]{hyperref}
\usepackage{booktabs}
\usepackage{tabularx}   
\usepackage{booktabs}   
\usepackage{array} 
\usepackage{pgffor}    
\usepackage{comment}
\usepackage{balance}

\usepackage{float}
\begin{document}

\title{Symmetry-Controlled Criticality and Zeno Localization in a Monitored Quasiperiodic Superconductor}

\author{Pinaki Singha}
\affiliation{Department of Physics, Indian Institute of Science Education and Research, Bhopal, Madhya Pradesh 462066, India}

\author{Marcin Szyniszewski}
\affiliation{Department of Computer Science, University of Oxford, Parks Road, Oxford OX1 3QD, UK}

\author{Nilanjan Roy}
\affiliation{Department of Physics, School of Advanced Sciences, VIT-AP University, Amaravati, Andhra Pradesh 522241, India}

\author{Auditya Sharma}
\affiliation{Department of Physics, Indian Institute of Science Education and Research, Bhopal, Madhya Pradesh 462066, India}

\date{\today}

\begin{abstract}
    Measurement-induced transitions of free fermions depend strongly on symmetry, but their interplay with quasiperiodic localization and superconducting pairing remains largely unexplored. We study a continuously monitored model of free fermions with pairing, where quasiperiodicity both controls localization and changes the symmetry class of the monitored field theory. Using topological entanglement entropy as a probe of nonlocal correlations in quantum state diffusion simulations, we identify a transition between a critical phase whose von Neumann entanglement entropy scales as $(\ln L)^2$ and an area-law phase, whereas no transition occurs in the clean limit, consistent with the corresponding DIII and D nonlinear sigma models. The transition line exhibits strong multifractality and depends nonmonotonically on quasiperiodicity, which we associate with the symmetry crossover. Deep in the Zeno regime, we derive an effective non-Hermitian Bogoliubov-de Gennes description of individual trajectories and obtain a closed-form localization length that quantitatively agrees with numerics. Our results reveal how symmetry, quasiperiodic localization, and measurement backaction combine to determine both critical entanglement and microscopic trajectory localization in monitored superconductors.
\end{abstract}
\maketitle

\let\oldaddcontentsline\addcontentsline
\renewcommand{\addcontentsline}[3]{}

\emph{Introduction.---}%
Competition between entangling unitary dynamics and local measurements can drive a sharp change in the entanglement structure of quantum states~\cite{Skinner2019, Chan2019, Li2018}. As the measurement rate is varied, monitored systems undergo measurement-induced phase transitions (MIPTs) between volume-law (extensive) and area-law entangled phases; MIPTs have been established theoretically in hybrid quantum circuits and monitored Hamiltonian evolution~\cite{Skinner2019, Chan2019, Li2018, Szyniszewski2019, Li2019, Szyniszewski2020, Bao2020, Jian2020, Zabalo2020, Gullans2020, Choi2020, Gullans2020scalable, Turkeshi2020, Bao2021, Nahum2021, Ippoliti2021, Lunt2021, Lavasani2021, Li2021, Potter2022, Sharma2022, Sierant2022, Boorman2022, Zabalo2022, Fisher2023, Dehghani2023, DiFresco2024metrology, Ha2024, Wang2026}, and have been observed experimentally in trapped-ion and superconducting-qubit platforms~\cite{Noel2022, Koh2023, Hoke2023}.
Noninteracting fermions provide a particularly interesting setting for measurement-induced criticality~\cite{Cao2019, Chen2020, Alberton2021, Buchhold2021, Turkeshi2021, Coppola2022, Carollo2022, Jian2022, Turkeshi2022, Szyniszewski2023, Merritt2023, Loio2023, Jian2023, Kells2023, Poboiko2024, Chahine2024, Lumia2024, Starchl2025, Chahine2025, Fava2023, Singha2026, Poboiko2023, Fava2024, Niederegger2026, Poboiko2025, Poboiko2026}: for short-range Gaussian dynamics, monitoring destabilizes volume-law entanglement~\cite{Cao2019, Fidkowski2021howdynamicalquantum}, allowing transitions between subextensively entangled phases.
Gaussianity enables efficient simulation, but finite-size effects can be exceptionally strong. A paradigmatic example is the one-dimensional, particle-number-conserving tight-binding chain, where numerics suggest an extended regime with logarithmic entanglement~\cite{Alberton2021}, whereas nonlinear sigma-model (NLSM) analyses and large-scale simulations find eventual saturation of entropy at any nonzero measurement strength~\cite{Poboiko2023, Fava2024, Fan2026, Niederegger2026, Yin2026}.

\begin{figure}[b!]
    \centering
    \includegraphics[width=\columnwidth]{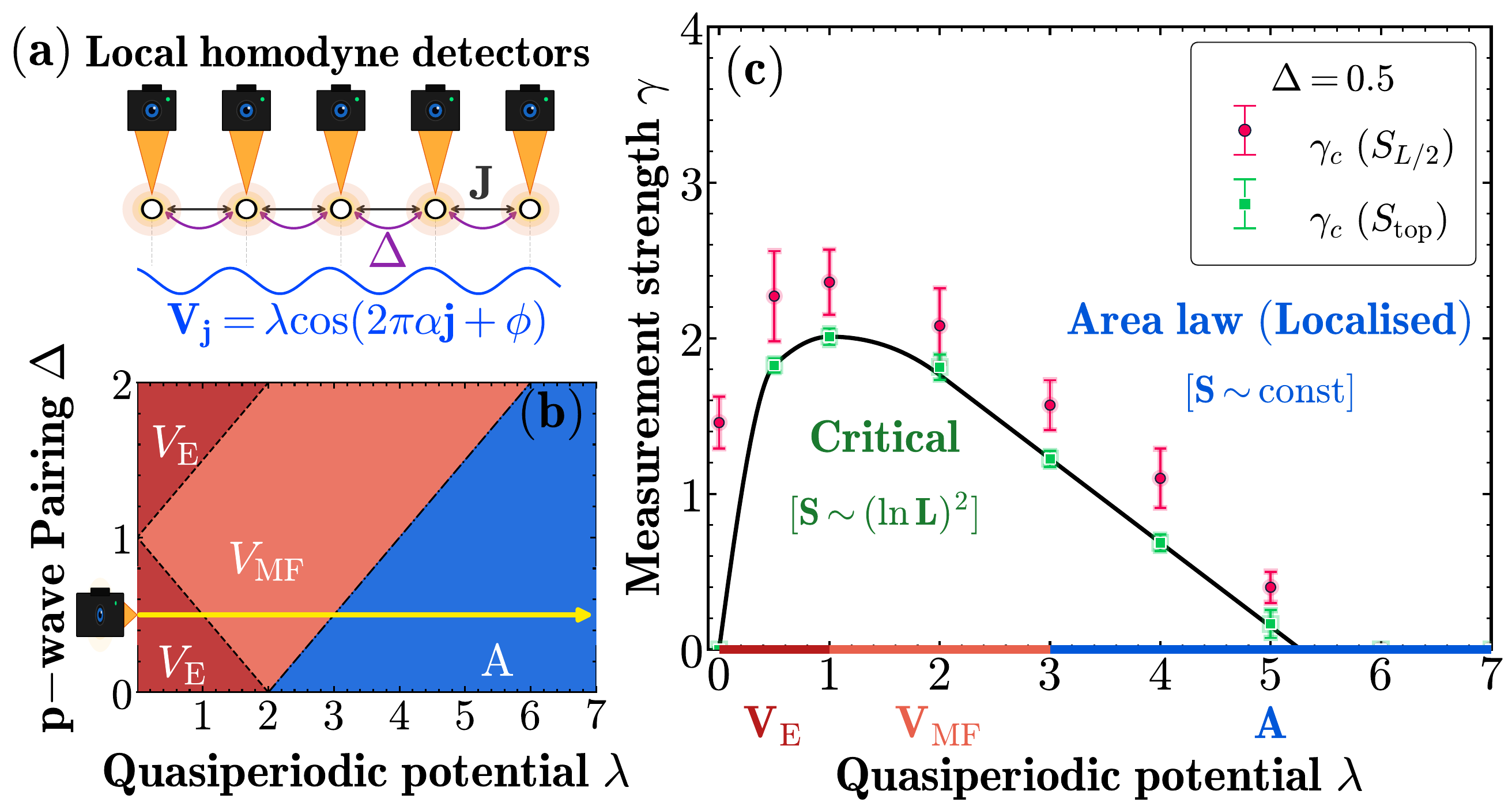}
    \caption{\textbf{(a)}~Monitored free fermions with hopping $J$, pairing $\Delta$, quasiperiodic potential $V_j$, and continuous local homodyne measurements of strength $\gamma$. \textbf{(b)}~Unmonitored phase diagram~\cite{Zeng_2018} showing volume-law extended ($\rm V_E$), volume-law multifractal ($\rm V_{MF}$) , and area-law localized (A) phases; we probe $\Delta=0.5$ (yellow arrow). 
    \textbf{(c)}~Phase diagram showing a measurement-induced phase transition between a critical $(\ln L)^2$-entangled phase and area law phase, estimated using entanglement entropy $S_{L/2}$ and topological entropy $S_{\rm top}$.}
    \label{fig:phase}
\end{figure}

The symmetry of the effective NLSM provides a framework for classifying measurement-induced phases and transitions in free fermions~\cite{Fava2023, Xiao2025, Bhuiyan2025}. Generic monitored Majorana chains without particle-number conservation realize a class DIII NLSM~\cite{Fava2023, Jian2023}, related to that of Anderson localization but in a different replica limit. Its stable weak-monitoring phase is analogous to a thermal metal and exhibits entanglement scaling as $S\sim(\ln L)^2$. The transition to area law is governed by a critical point with $S\sim\ln L$ and a finite correlation-length exponent~\cite{Fava2023}. Additional symmetries can instead yield a class D NLSM without a stable metallic phase in the monitored replica limit~\cite{Fava2023}, despite pronounced finite-size crossovers~\cite{Turkeshi2021}. Understanding how monitored dynamics crosses between these classes, and how to characterize the resulting phases beyond entanglement scaling, is of active research interest.

Particle-number-nonconserving systems are also appealing topologically: $p$-wave pairing is the defining ingredient of the Kitaev chain, whose topological phase hosts Majorana boundary modes~\cite{Kitaev2001, Mourik2012}, while monitoring competing observables can reshape topological structure~\cite{Lavasani2021, Kells2023}. Quasiperiodicity provides another independently tunable ingredient, producing an analytically tractable localization transition even in one dimension, as seen in the Aubry-Andr\'e-Harper (AAH) model~\cite{Harper1955, AubryAndre1980}. In $p$-wave superconductors, it strongly modifies the spatial structure and stability of Majorana edge modes~\cite{Wang2016, Wang2016spectral, Zeng2016, Ghadimi2017, Zeng_2018, Fraxanet2021, Cheng2022}. In monitored superconductors, it therefore allows controlled departure from the symmetry-enhanced clean limit while introducing nontrivial spatial structure, thus providing a natural setting for studying symmetry, measurement-induced criticality, localization, and Majorana physics.

In this Letter, we exploit quasiperiodicity to tune a monitored superconducting model between class D at the clean point and class DIII at finite potential. We identify a robust transition diagnostic that reveals a multifractal transition line between the $(\ln L)^2$ critical phase and an area-law phase. The corresponding critical monitoring strength depends nonmonotonically on the potential and changes sharply near the clean limit, which we associate with the change in symmetry. We also study the area-law phase by developing a non-Hermitian effective theory yielding a closed-form localization length in terms of model parameters. Our results show how the symmetry classification of the monitored dynamics is reflected directly in the phase boundary and how quasiperiodicity modifies the spatial structure of trajectory states.


\emph{Model and method.---}%
We consider an open chain of $L$ spinless fermions~\cite{Wang2016, Wang2016spectral, Zeng_2018},
\begin{align}
    H &= - J \sum_{j=1}^{L-1}\big(c_{j+1}^\dagger c_j+\mathrm{H.c.}\big) + \Delta \sum_{j=1}^{L-1}\big(c_{j+1}^\dagger c_j^\dagger+\mathrm{H.c.}\big) \nonumber\\
    &\quad +\sum_{j=1}^{L} V_j c_j^\dagger c_j, \qquad V_j=\lambda\cos(2\pi\alpha j+\phi).
    \label{eq:H}
\end{align}
The hopping $J$ delocalizes particles, the superconducting pairing $\Delta$ hybridizes particles and holes, and the potential strength $\lambda$ controls the quasiperiodic localization. We take $\alpha=(\sqrt5-1)/2$ and a random offset $\phi$ for each quantum trajectory. The model reduces to the AAH model at $\Delta=0$ (with a localization transition at $\lambda = 2J$), and to the Kitaev chain at $\lambda = 0$, where the system separates into two Majorana subchains coupled by measurements, yielding an enhanced symmetry class D; onsite potentials break this, restoring the generic class DIII.
The unmonitored Bogoliubov modes are extended for $\lambda < 2|J - \Delta|$, multifractal for $2|J - \Delta| < \lambda < 2(J + \Delta)$, and exponentially localized for $\lambda > 2(J + \Delta)$ [see Fig.~\ref{fig:phase}(b)]~\cite{Wang2016, Wang2016spectral, Zeng_2018}; the first two regimes produce volume-law entanglement, while the localized regime is area-law entangled. We fix $J = 1$, $\Delta = 0.5$, and vary $\lambda$ to probe measurement effects across these regimes.

We continuously monitor every local occupation $n_j = c_j^\dagger c_j$ using quantum state diffusion (QSD) [see Fig.~\ref{fig:phase}(a)], described by the stochastic Schr\"odinger equation,
\begin{align}
    d|\psi\rangle=&\Big[-iH-\frac{\gamma}{2}\sum_j(n_j-\langle n_j\rangle)^2\Big]dt\,|\psi\rangle\notag\\
    &+\sum_j\,(n_j-\langle n_j\rangle)\,dW_j\,|\psi\rangle,
    \label{eq:qsd}
\end{align}
where $\gamma$ is the measurement strength and $dW_j$ is the It\^o increment with zero mean and variance $\gamma \, dt$. The dynamics conserves fermion parity but not particle number. Each trajectory remains Gaussian: $H$ is quadratic and $n_j^2=n_j$ makes every measurement update an exponential of a fermion bilinear. The state is therefore fully characterized by its Nambu correlation matrix, permitting large-system simulations~\cite{Mbeng2024}.
\begin{figure}[t]
    \centering
    \includegraphics[width=\columnwidth]{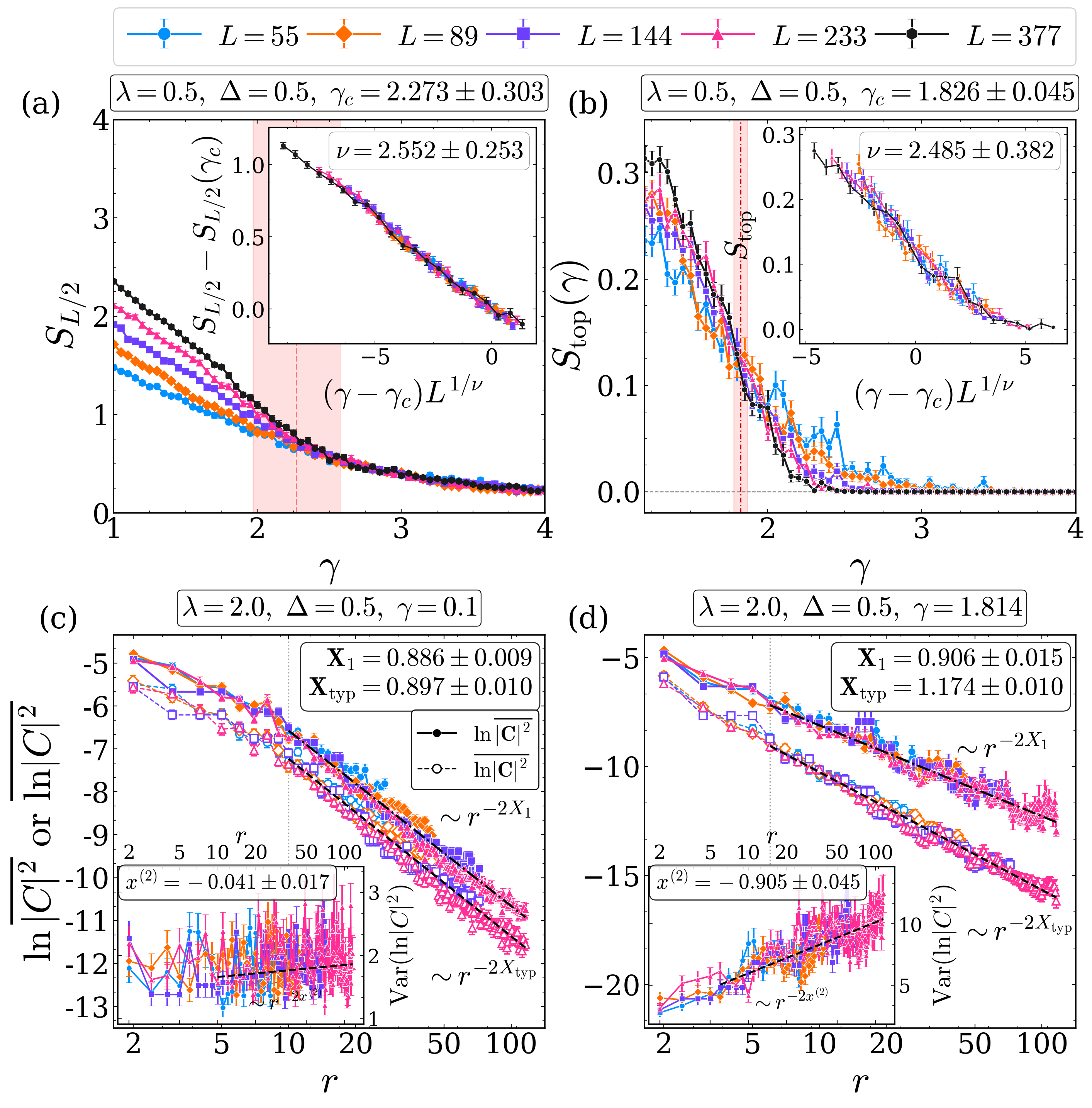}
    \caption{(a)~$S_{L/2}$ and (b)~$S_\mathrm{top}$ vs.\@ measurement rate $\gamma$ for $\lambda=0.5$ and $\Delta=0.5$. Insets show finite-size scaling collapse; the dashed red vertical line marks the extracted critical point $\gamma_c$, and the shaded region indicates the uncertainty $\pm\gamma_{c,\mathrm{err}}$. (c,d)~$\ln\overline{|C(r)|^2}$ (filled symbols) and $\overline{\ln|C(r)|^2}$ (open symbols) as a function of $r$, for $\lambda=2$, $\Delta=0.5$ at (c)~$\gamma=0.1$ (below transition point) and (d)~$\gamma=1.814\approx\gamma_c$ (at the transition point).
    Black lines are power-law fits. Insets show the variance of $\ln|C(r)|^2$.}
    \label{fig:c_Stop}
\end{figure}
Numerically, we use an efficient representation of the state by a $2L\times L$ matrix of occupied modes $\mathcal U = (U, V)^T$~\cite{Turkeshi2021, Mbeng2024}. We start the evolution from a vacuum state ($U = \mathbb{1}, V = \mathbb{0}$), and evolve it using Trotterized steps of Eq.~\eqref{eq:qsd}, $\mathcal U_{t+dt} = \mathcal{N} e^{2M} e^{-2 i dt H_{\rm BdG}} \mathcal U_{t}$, where $H_{\rm BdG}$ is the $2L \times 2L$ Nambu form of the Hamiltonian~\eqref{eq:H}, and similarly $M_j = (dW_j + (2\langle n_j\rangle-1) \gamma dt) n_j$ is the measurement written in the Nambu form. Normalization $\mathcal N$ is performed using QR decomposition, which also restores orthonormality of $\mathcal U$. We find the time discretization of $dt=0.05$ sufficient for convergence, and probe the steady state at time $T=1200$. We choose Fibonacci system sizes to reduce finite-size artifacts and improve convergence~\cite{Cestari2010, Yi2025, Hetenyi2025, Yildiz2026}. All observables follow from the correlation matrix $\mathcal{G} = \mathcal{U} \mathcal{U}^\dagger$.


\emph{Measurement-induced entanglement transition.---}%
A natural transition diagnostic is the von Neumann entanglement entropy $S_A$, defined for a bipartition into subsystem $A$ and its complement as $S_A = -\sum_m \zeta_m \ln \zeta_m$, where $\{\zeta_m\}$ is the full spectrum of the correlation matrix $\mathcal{G}_A$ restricted to subsystem $A$. The steady-state half-chain entropy $S_{L/2}$ grows with $L$ at weak monitoring and saturates for sufficiently large monitoring strengths [see Fig.~\ref{fig:c_Stop}(a)]. Our data are consistent with the DIII prediction $S\sim(\ln L)^2$ in the critical phase~\cite{Fava2023}. At the transition line, the RG flow terminates at a scale-invariant critical fixed point with $S \sim \ln L$, which is hard to distinguish numerically and requires finite-size scaling analysis.

We also inspect the topological entanglement entropy,
\begin{equation}
    S_{\rm top} = S_{AB} + S_{BD} - S_B - S_{ABD},
\end{equation}
for a partition $ABCD$ (each subsystem of size $L/4$). This combination cancels local boundary terms in $S$, and can be interpreted as conditional mutual information $I(A:D|B)$ between regions $A$ and $D$, given region $B$. We retain the terminology of Ref.~\cite{Kells2023}, where it diagnosed topological phases, but use it here to probe nonlocal correlations. In the area law phase, $S_{\rm top}$ is expected to decay as $e^{-|C|/\xi}$, where $|C|$ is the size of region $C$, and $\xi$ is the correlation length, while in the critical phase it is expected to grow $\sim \ln L$ due to subleading $\ln L$ entropy contributions [see Fig.~\ref{fig:c_Stop}(b)].

Hence, we use the one-parameter scaling forms,
\begin{align}
    S_{L/2}(\gamma,L)-S_{L/2}(\gamma_c,L)=F[(\gamma-\gamma_c)L^{1/\nu}],
    \label{eq:S_collapse}\\
    S_{\rm top}(\gamma,L)=G[(\gamma-\gamma_c)L^{1/\nu}],
    \label{eq:Stop_collapse}
\end{align}
where $F[\cdot], G[\cdot]$ are universal scaling functions and $\nu$ is the correlation length exponent. Repeating the data collapse [Fig.~\ref{fig:c_Stop}(a,b), insets] 
across different values of $\lambda$ yields the phase boundary in Fig.~\ref{fig:phase}(c) (see End Matter for all $S_{\rm top}$ collapses and Supplemental Material (SM) for further details~\cite{SuppMat}). Estimates of the transition point from $S_{\rm top}$ are systematically lower than those from $S_{L/2}$. At $\lambda=0$, $S_{\rm top}$ exhibits no size-dependent crossings, consistent with the absence of a transition in the class D description, whereas $S_{L/2}$ predicts an apparent finite $\gamma_c$. This discrepancy indicates stronger finite-size effects in $S_{L/2}$ and motivates using $S_{\rm top}$ as the primary diagnostic.

The monitored critical phase persists beyond the unmonitored localization threshold, and the underlying extended, multifractal, and localized regimes leave little visible imprint on the phase boundary at $\gamma>0$. Likely, measurement backaction continually reshapes the Gaussian orbitals, partially counteracting coherent localization. Furthermore, the correlation-length exponents extracted from $S_{\rm top}$ are consistent, within uncertainties, with an approximately constant value $\nu \simeq 1.9$ along the transition line, with no statistically significant dependence on $\lambda$, compatible with a common universality class across the finite-$\lambda$ DIII transition line. The corresponding $S_{L/2}$ estimates are likewise approximately constant, but systematically larger, $\nu \simeq 2.6$, consistent with their stronger finite-size corrections.

The phase boundary is strongly nonmonotonic: weak quasiperiodicity $\lambda \sim J$ initially increases the monitoring strength required to destroy the critical phase, while stronger potential eventually suppresses it completely beyond $\lambda \approx 5.3$. We associate the sharp change near the clean limit with the D-to-DIII symmetry change. Analogous nonmonotonicity in the U(1)-symmetric model~\cite{Szyniszewski2023} has been attributed to a change in preasymptotic entanglement dynamics between classes AIII and BDI~\cite{Mantik2026inpreparation}. Similar nonmonotonicity was observed numerically in monitored parity-preserving free fermions with random disorder~\cite{Szyniszewski2024}, yet neither its connection to symmetry nor the absence of a clean-limit MIPT was established. Our results suggest that here the nonmonotonicity survives at the thermodynamic phase boundary.

We further probe the distribution of correlations through the moments of $C(r)=\langle c^\dagger_{(L-r)/2} c_{(L+r)/2} \rangle$, which scale as $\overline{|C(r)|^{2n}} \sim r^{-2X_n}$. Following Ref.~\cite{Jian2023}, multifractality can be diagnosed from the average exponent $X_1$, the typical exponent $X_{\rm typ}$ extracted from $\overline{\ln |C(r)|^2}$, and the exponent $x^{(2)}$ governing the variance of $\ln |C(r)|^2$. At the transition line [see Fig.~\ref{fig:c_Stop}(d)] we find $X_1 < X_{\rm typ}$ and $x^{(2)}<0$, showing that average and typical correlations decay with different exponents and that the distribution of $\ln |C(r)|^2$ broadens with distance. This reveals a broad, multifractal distribution of correlations and a nonlinear spectrum of $X_n$. This behavior disappears inside the critical phase [see Fig.~\ref{fig:c_Stop}(c)], where $X_1 \simeq X_{\rm typ}$ and $x^{(2)} \simeq 0$, showing no evidence of conventional multifractality, likely due to the marginal flow of the DIII NLSM in this phase. We cannot exclude parametrically weaker multifractality, as the analogy with weakly disordered metals would suggest~\cite{Mirlin2000}.


\emph{Zeno localization through the effective non-Hermitian description.---}%
The entanglement transition identifies the onset of the area-law phase but does not determine its microscopic localization scale. Deep in the quantum-Zeno regime, we propose an effective non-Hermitian Bogoliubov description of the spatial decay of trajectory orbitals, extending the particle-number-conserving construction of Ref.~\cite{Singha2026} to paired Gaussian states; the coupled particle and hole sectors require a substantial modification. Detailed derivations are given in SM~\cite{SuppMat}.

\begin{figure}[t]
    \centering
    \includegraphics[width=\columnwidth]{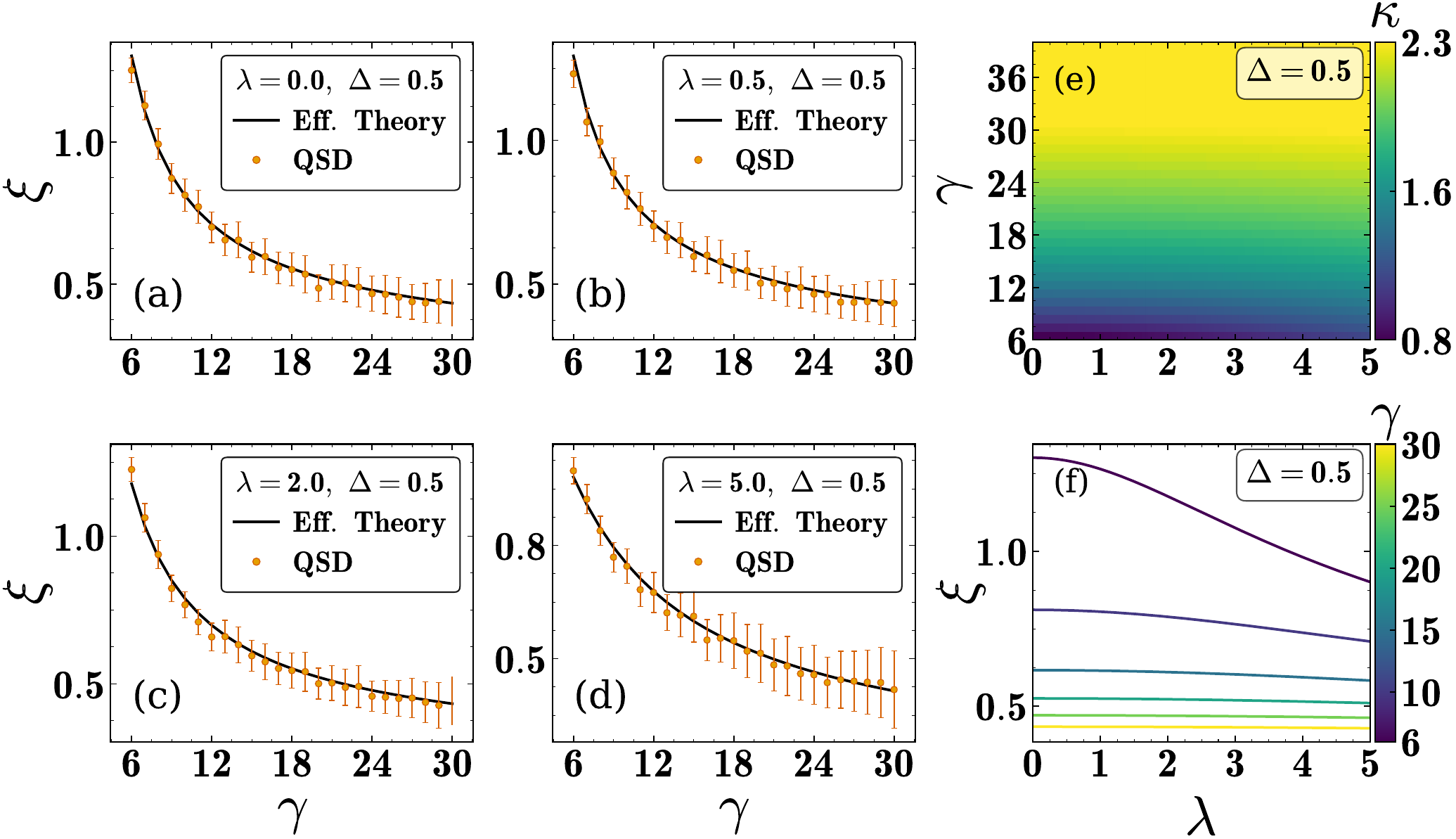}
    \caption{Localization length $\xi$ vs.\ $\gamma$ for (a)~$\lambda=0.0$, (b)~$\lambda=0.5$, (c)~$\lambda=2.0$, and (d)~$\lambda=5.0$, at $\Delta=0.5$, $J=1$, $L=233$. Solid curves: effective theory [Eq.~\eqref{eq:xi_main}]; symbols: QSD numerics. (e)~Lyapunov exponent $\kappa=1/\xi$ vs.\ $\gamma$ and $\lambda$. (f)~Correlation length $\xi$ vs.\ $\lambda$. }
    \label{fig:xi}
\end{figure}

Strong measurements suppress hopping and separate time scales. We observe concentration of the energy density near zero, which motivates a zero-energy reference description of the orbital tails (see End Matter). The trajectories relax onto a manifold of exponentially localized pointer states: an instantaneous Nambu spinor $\Psi_j = (u_j,v_j)^T$ can be written as a reference state $\Phi_0(j)$ (localized over $\mathcal O(1)$ sites) plus fluctuations.
$\Phi_0(j)$ can be described using the stationary effective equations.
For a reference orbital approaching a site-localized particle or hole state, orthogonality gives $\mathcal{O}(\epsilon)$ fluctuations at its dominant core component,
where $\epsilon = {\sqrt{J^2 + \Delta^2}} / {\sqrt{\lambda^2 + (\gamma/2)^2}} \ll 1$ is a model-dependent parameter (see End Matter).
At the reference energy $E_{\text{dom}} = 0$, we define the reconstructed
potential $V_j^{\rm QSD}(t)$, wherever $u_j\neq 0$, through
\begin{equation}
    -J(u_{j+1}{+}u_{j-1}) + V_j^{\rm QSD}(t)u_j + \Delta(v_{j-1}{-}v_{j+1}) = E_{\rm dom}u_j
\end{equation}
(and likewise for the hole component $v_j$); expanding about the stationary reference profile gives
\begin{equation}
    V_j^{\rm QSD}(t) = V_j - i\frac{\gamma}{2} + i\gamma \langle n_j\rangle_0 + \delta V_j^{\rm stoch}(t) ,
    \label{eq:Vqsd_full_main}
\end{equation}
where $\delta V_j^{\rm stoch}(t)$ is a stochastic correction.
The core contribution $i\gamma\langle n_j\rangle_0$ decays exponentially away from the localization center, and therefore does not change the asymptotic decay rate.
Under simple assumptions on the reference state (which we verify numerically), the reconstructed potential has a relative correction $\mathcal{O}(\epsilon)$, and the QSD orbital shares the asymptotic decay exponent of the reference, its tail obeying an effective Hamiltonian
\begin{equation}
    H_{\rm eff} = H_{\rm BdG} - i\frac{\gamma}{2}\,\tau_z,
    \label{eq:Heff_main}
\end{equation}
where $\tau_z = \mathrm{diag}(\mathbb{1}_L, -\mathbb{1}_L)$ acts in Nambu (particle-hole) space. This represents the measurement drift by a uniform imaginary onsite potential in Nambu space.

In a Majorana basis, the corresponding transfer-matrix recursion is $\psi_{j+1}=a_j\psi_j+b\,\psi_{j-1}$ with $a_j=(V_j-i\gamma/2)/(J+\Delta)$, $b=-(J-\Delta)/(J+\Delta)$. Thus, superconducting pairing renormalizes both forward and backward propagation in comparison to the $\Delta=0$ case. Since $|a_j|=\mathcal O(\epsilon^{-1})\gg|b|=\mathcal O(1)$ in the Zeno regime, backward coupling is perturbatively suppressed, and the transfer-matrix product defining the Lyapunov exponent becomes effectively multiplicative, $\kappa \approx \overline{\ln|a_j|} + \mathcal O(\epsilon^2)$, yielding the closed form for the localization length
\begin{equation}
    \xi = \left[\ln\!\left(
    \frac{\sqrt{\lambda^2 + (\gamma/2)^2} + \gamma/2}{2(J + \Delta)}
    \right)\right]^{-1}
    + \mathcal{O}(\epsilon^2),
    \label{eq:xi_main}
\end{equation}
asymptotically exact in the Zeno limit, and reducing to the monitored-AAH result at $\Delta=0$~\cite{Singha2026} (see End Matter for further comments).
Crucially, this effective theory can be adapted to Anderson disorder and other static onsite potentials~\cite{Sau2012, fortin2026finitedisorderedkitaevchains, PhysRevA.84.013612, PhysRevB.103.224505}, providing a general framework for Zeno-localized superconducting chains without particle number conservation.

We compare Eq.~\eqref{eq:xi_main} from the effective theory with QSD simulations, where we used $dt=0.01$ to account for larger $\gamma$. Following the method of Ref.~\cite{Szyniszewski2024}, we reconstruct localized Bogoliubov orbitals and fit their amplitude tails to $e^{-r/\xi}$. The numerics in Fig.~\ref{fig:xi}(a--d) spans clean, extended, multifractal, and localized unmonitored regimes and agrees quantitatively with the effective theory to $\mathcal{O}(\epsilon^2)$. Fig.~\ref{fig:xi}(e,f) maps the predicted Lyapunov exponent and localization length across the space of $(\lambda, \gamma)$. These results clarify the relation between quasiperiodicity and measurements: near the entanglement transition, quasiperiodicity can stabilize the critical phase, but deep in the Zeno regime, both quasiperiodicity and monitoring increase the decay rate and cooperate to localize the single-particle orbitals.


\emph{Discussion.---}%
Our results show how symmetry impacts measurement-induced transitions in the monitored quasiperiodic superconductor. Finite quasiperiodicity realizes the DIII monitored field theory, yielding a genuine transition between a stable $(\ln L)^2$-entangled phase and an area-law phase. The transition exhibits a common universal behavior and hosts strongly multifractal correlations across the phase boundary. The clean limit exhibits no transition, consistent with the class D symmetry prediction. We associate the nonmonotonic behavior of the phase boundary and severe finite-size effects near this limit with the DIII-to-D symmetry crossover. We specifically find that a nonlocal entanglement measure such as the topological entanglement entropy resolves this behavior reliably even near the D-class point. Within the area-law phase, we have developed a Zeno-limit effective theory of particle-nonconserving Gaussian states, linking trajectory localization to a non-Hermitian BdG problem. This gives a closed-form localization length that agrees with numerics and recovers the particle-conserving limit, revealing an interplay of quasiperiodicity and monitoring on spatial wave function properties. Importantly, this effective theory can be adapted to other monitored free fermion models.

Several directions follow naturally from these results. A field-theoretic treatment of critical-phase correlations could determine whether weaker multifractality survives beyond our numerical resolution. Combining quasiperiodicity with competing measurements~\cite{Kells2023} or competing interactions~\cite{Xu2023} could reveal topological counterparts of the symmetry-controlled transition through Majorana edge modes, entanglement-spectrum degeneracies, and trajectory topological invariants. In this context, the non-Hermitian BdG and Lyapunov description developed here provides a starting point for relating Zeno localization to bulk-edge correspondence in monitored superconductors. Extending these ideas to higher-winding-number Kitaev chains and two-dimensional $p+ip$ superconductors would test its scope, while circuit-QED and nanowire platforms with local charge measurements offer promising experimental routes. More broadly, our results identify monitored quasiperiodic superconductors as a setting in which genuine measurement-induced criticality, Anderson-localization physics, topology, and non-Hermitian trajectory dynamics can be explored within a common framework, while opening a route toward understanding how measurement constraints affect the stability of topological quantum information.


\emph{Acknowledgments.---}%
The authors would like to thank Antonio Garc\'ia-Garc\'ia, Lukas Haller, Xin-Chi Zhou, Aditi Chakrabarty, Igor Poboiko, and Igor Gornyi for useful comments. P.~S.\ acknowledges IISER Bhopal for the Ph.D.\ fellowship. M.~S.\@ was supported by the EPSRC grant on (De)constructing Quantum Software (DeQS) (Grant Reference No. EP/Z002230/1). We thank the HPC facility at IISER Bhopal, the University of Oxford Advanced Research Computing (ARC) facility, and the PARAM Rudra supercomputing facility at the Inter-University Accelerator Centre (IUAC), New Delhi, for computational resources.

\emph{Data availability.---}%
The data and code that support the findings of this Letter are openly available at Ref.~\cite{data_repository}.

\bibliography{refs}


\onecolumngrid
\begin{center}
{\large\bfseries End Matter}
\end{center}
\vspace{4pt}
\twocolumngrid


\twocolumngrid
\begin{figure*}[t]
    \centering
    \includegraphics[width=\textwidth]{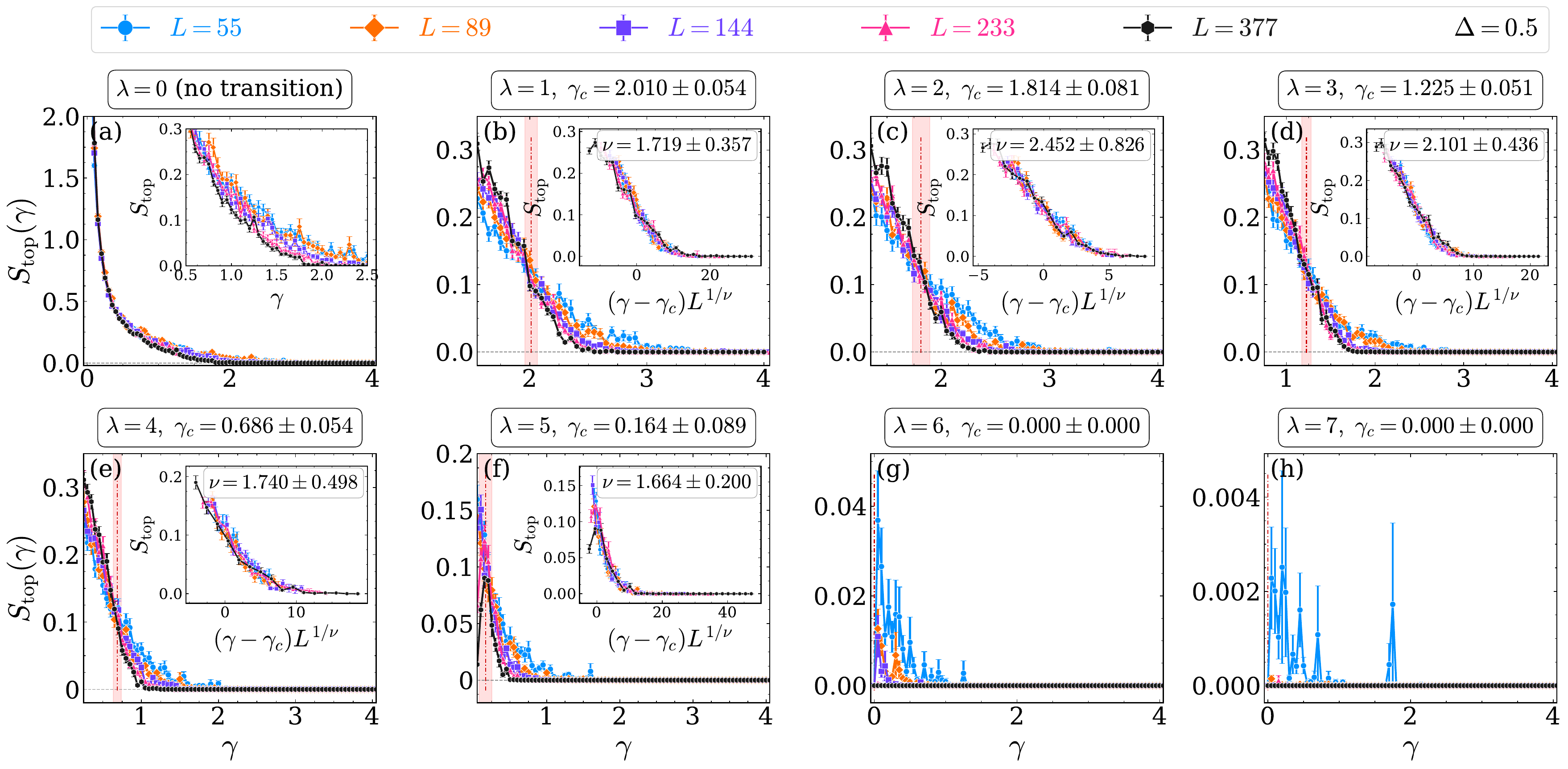}
    \caption{Topological entanglement entropy $S_{\rm top}(\gamma)$ for $\Delta=0.5$, and $\lambda=0$--$7$. Dash-dotted line and shaded band mark the critical point and its statistical uncertainty. The insets show data collapses to the finite-size scaling ansatz in Eq.~\eqref{eq:Stop_collapse} with extracted collapse parameters annotated. Panel~(a) ($\lambda=0$) shows no crossing, consistent with the class-D limit. 
    Panels~(g)--(h) ($\lambda=6,7$) have $\gamma_c=0$ and are shown without an inset.}
    \label{fig:Stop_grid}
\end{figure*}

\emph{Finite-size scaling of topological entanglement entropy.---}%
Here we provide supplementary results for topological entanglement entropy and its finite-size scaling collapse, shown in Fig.~\ref{fig:Stop_grid}.
For $\lambda=0$ the chain decouples into two class-D Majorana chains with no phase transition, consistent with the absence of any size-dependent crossing in panel~(a). 
Turning on $\lambda>0$ changes the symmetry class to DIII, visible as a presence of crossing of $S_{\rm top}(\gamma)$ curves for different $L$.

Weighted constant fit to the extracted correlation length critical exponents gives $\nu \approx 1.86\pm 0.14$, with reduced chi-squared of $\chi^2/(N-1) \simeq 0.934$. So there is no statistically significant evidence for variation of $\nu$ with $\lambda$, which is compatible with a common universality class across the transition line.

\emph{Dominant energy scale of the monitored Bogoliubov--de Gennes chain.}---%
Particle--hole symmetry fixes the average of instantaneous energy to zero and self-averaging collapses the distribution onto it, pinning the dominant energy to zero in the thermodynamic limit.

\begin{figure*}[t]
\centering
\includegraphics[width=\textwidth]{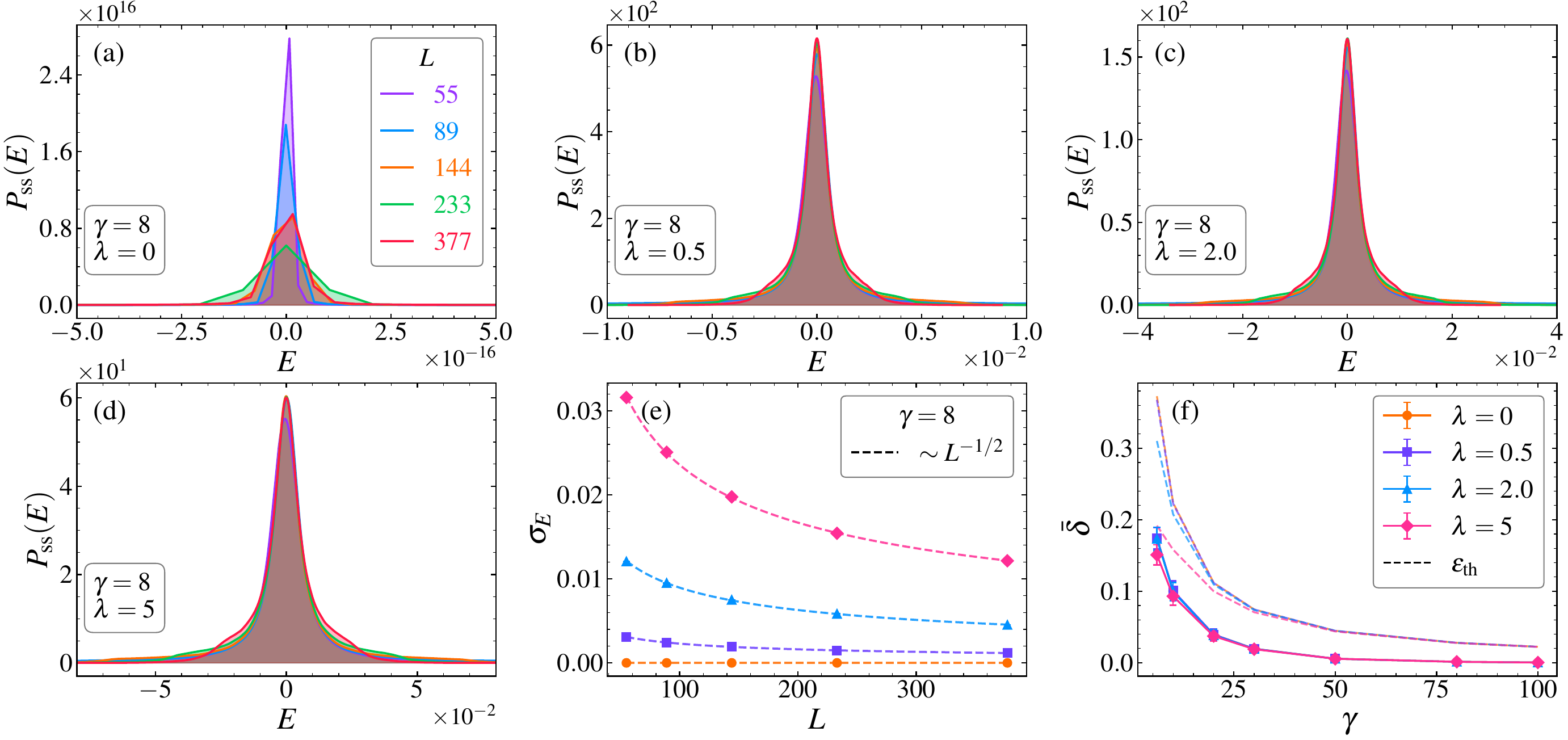}
\caption{Steady-state distributions $P_{\rm ss}(E)$ at $\gamma=8$
for (a) $\lambda=0$, (b) $0.5$, (c) $2.0$, (d) $5.0$
with $L=55,89,144,233,377$.
(e) $\sigma_E$ vs.\ $L$ showing $\sigma_E\sim L^{-1/2}$ (dashed).
(f) Mean localization floor $\bar{\delta}$ vs.\ $\gamma$; dashed
lines show
$\varepsilon_{\rm th}=\sqrt{J^2+\Delta^2}/\sqrt{\lambda^2+(\gamma/2)^2}$.}
\label{fig:combined}
\end{figure*}

The instantaneous energy density $E_{\mathrm{inst}}(t)=\frac{1}{2N}\mathrm{Tr}[H_{\mathrm{BdG}}\mathcal{G}(t)]$ has stationary distribution $P_{\mathrm{ss}}(E) = \lim_{T\to\infty} \frac{1}{T} \int_{t_0}^{t_0+T} \delta(E - E_{\mathrm{inst}}(t)) \, dt$, peaking at the dominant energy $E_{\mathrm{dom}} \simeq 0$ for all $\lambda$ [see Fig.~\ref{fig:combined}(a)--(d)].
Particle--hole symmetry enforces $P_{\mathrm{ss}}(E) = P_{\mathrm{ss}}(-E)$, hence $\langle E_{\mathrm{inst}}\rangle=0$. In the Zeno regime, $\mathrm{Cov}(\epsilon_i,\epsilon_j) \sim e^{-|i-j|/\xi}$ for two energy levels $\epsilon_i, \epsilon_j$, which gives $\sigma_E\sim L^{-1/2}$ [see Fig.~\ref{fig:combined}(e)], so $\lim_{L \to \infty} P_{\mathrm{ss}}(E) = \delta(E)$ and $E_{\mathrm{dom}} = 0$ independent of $\lambda, \gamma$.

\emph{Numerical check of the bound on the fluctuation field.}---%
We verify that the localized Bogoliubov quasiparticle wavefunctions obtained from QSD satisfy the obtained $\mathcal{O}(\epsilon)$ bound (see SM~\cite{SuppMat})
\begin{equation}
  \overline{\delta} \lesssim \varepsilon = \frac{\sqrt{J^2 + \Delta^2}}{\sqrt{\lambda^2 + (\gamma/2)^2}},
  \label{eq:bound_bdg}
\end{equation}
where $\varepsilon \ll 1$ controls the Zeno hierarchy. For each trajectory, the QSD wavefunction is unscrambled into maximally localized orbitals; for each orbital with localization center $j_0$,
\begin{equation}
  \delta^2 = \sum_{j \neq j_0}(|U_j|^2 + |V_j|^2) = 1 - (|U_{j_0}|^2 + |V_{j_0}|^2) \simeq \|\delta\psi\|^2,
\end{equation}
where $U_j, V_j$ are the particle and hole amplitudes of the Nambu spinor, and $\delta^2$ directly measures the norm-squared of the fluctuation field $\delta\psi$ away from the dominant localized component. We average $\delta^2$ over $L=233$ orbitals and $128$ independent trajectories and take the square root to obtain $\overline{\delta}$, with $T=1200$, $dt=0.05$, $J=1$, $\Delta=0.5$. As shown in Fig.~\ref{fig:combined}(f), $\overline{\delta}$ lies strictly below $\varepsilon(\gamma,\lambda)$ for all parameter sets; the gap grows with $\gamma$ and $\overline{\delta}\to 0$ in the Zeno regime, confirming that the fluctuation expansion becomes asymptotically exact as $\varepsilon\to 0$.

\emph{Relation to the particle-conserving theory.}---%
Eq.~\eqref{eq:xi_main} reduces to the monitored-AAH result of Ref.~\cite{Singha2026} as $\Delta \to 0$, providing a nontrivial consistency check. Although both approaches extract an effective non-Hermitian localization problem from typical QSD trajectories, pairing changes its structure fundamentally: the Slater-orbital dynamics of Ref.~\cite{Singha2026} is replaced by a constrained Nambu evolution of coupled particle and hole amplitudes. Additionally, particle-hole symmetry fixes the relevant zero-energy sector and enforces cancellations absent in the number-conserving problem. Moreover, the BdG transfer equation contains two distinct propagation channels, $\psi_{j+1} = \frac{V_j - i \gamma / 2}{J + \Delta} \psi_j - \frac{J - \Delta}{J + \Delta}\psi_{j-1}$, so that the scale $J+\Delta$ emerges only after the second channel is perturbatively suppressed in the Zeno limit. The resulting localization length therefore cannot be obtained from the particle-conserving theory by an \emph{a priori} renormalization of the hopping. The physical setting is also distinct: the model of Ref.~\cite{Singha2026} includes a smooth crossover to measurement-induced localization, whereas here the Zeno theory governs the area-law side of a genuine measurement-induced transition. Thus, the $\Delta \to 0$ limit connects the two theories smoothly, while finite pairing yields an independently constructed BdG generalization with qualitatively new symmetry and propagation structure.
\onecolumngrid
\input{supplementary}

\end{document}

%% file: supplementary.tex

\clearpage
\onecolumngrid
\begin{center}
\textbf{\large Supplementary Materials for:}\\
\textbf{Symmetry-Controlled Criticality and Zeno Localization in a Monitored Quasiperiodic Superconductor}
\end{center}
\begin{center}
Pinaki Singha,$^{1}$ Marcin Szyniszewski,$^{2}$ Nilanjan Roy,$^{3}$ and Auditya Sharma$^{1}$
\end{center}
\begin{center}
\small
$^{1}$Department of Physics, Indian Institute of Science Education and Research, 
Bhopal, Madhya Pradesh 462066, India\\
$^{2}$Department of Computer Science, University of Oxford, 
Parks Road, Oxford OX1 3QD, United Kingdom\\
$^{3}$Department of Physics, School of Advanced Sciences, VIT-AP University, 
Amaravati, Andhra Pradesh 522241, India
\end{center}


 
 
\renewcommand{\theequation}{S\arabic{equation}}
\renewcommand{\thefigure}{S\arabic{figure}}
\renewcommand{\thesection}{S\arabic{section}}
\setcounter{equation}{0}
\setcounter{figure}{0}



\setcounter{secnumdepth}{3}

\hypersetup{
    linktoc=all   
}

\makeatletter
\renewcommand{\l@section}{%
  \addvspace{1.5em}\@dottedtocline{1}{0em}{2.8em}%
}
\renewcommand{\l@subsection}{%
  \addvspace{0.5em}\@dottedtocline{2}{2.8em}{3.5em}%
}
\makeatother

\let\addcontentsline\oldaddcontentsline

\makeatletter
\def\l@subsubsection#1#2{}
\makeatother

\tableofcontents

\section{Finite-size scaling of the phase transition}
\label{sec:finitesize_scaling}

\begin{figure*}[t]
    \centering
    \includegraphics[width=\textwidth]{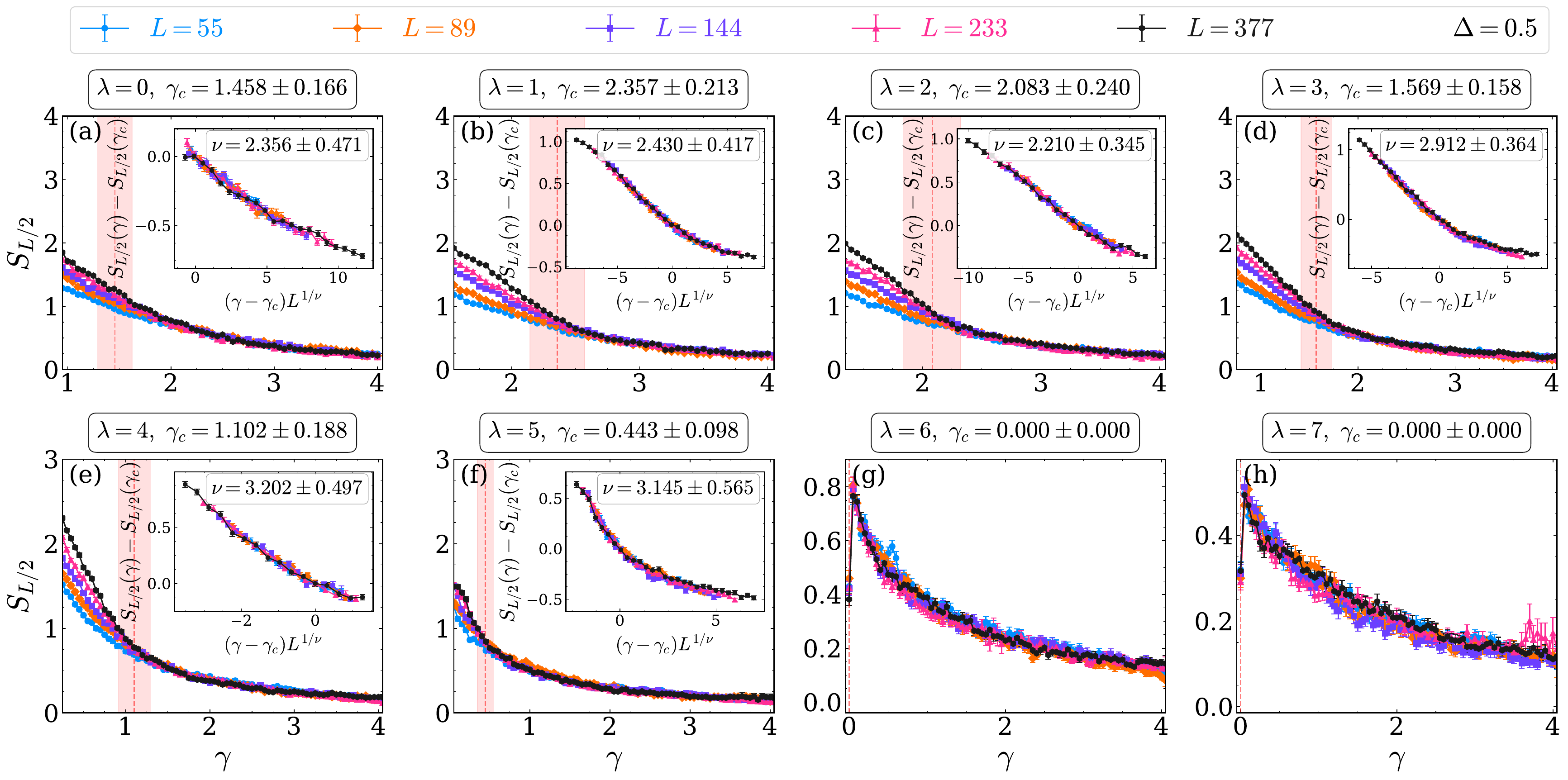}
    \caption{
    Finite-size scaling of the half-system entanglement entropy $S_{L/2}$ as a function of measurement strength $\gamma$ for all quasiperiodic potential strengths $\lambda$ studied.
    Each panel shows $S_{L/2}(\gamma)$ for different system sizes, with $\Delta=0.5$ and $J=1$.
    The dashed red vertical line marks the extracted critical point $\gamma_c$ and the shaded region indicates the uncertainty $\pm\gamma_{c,\mathrm{err}}$.
    Insets show the finite-size scaling collapse $S_{L/2}(\gamma) - S_{L/2}(\gamma_c)$ vs.\ $(\gamma-\gamma_c)L^{1/\nu}$ using the values of $\nu$ listed in Table~\ref{tab:S1}.
    Panels~(a)--(h) correspond to $\lambda = 0, 1, 2, 3, 4, 5, 6, 7$ respectively.
    }
    \label{fig:S1}
\end{figure*}

Table~\ref{tab:S1} summarizes the finite-size scaling results for the
measurement-induced phase transition, extracted from both
$S_\mathrm{top}$ and $S_{L/2}$ for all $\lambda$ values studied. Fig.~\ref{fig:S1} shows the finite-size scaling collapse of $S_{L/2}$ for all $\lambda$ values, while the collapses of $S_{\rm top}$ are shown in the End Matter of the main text.

\begin{table*}[b]
    \centering
    \renewcommand{\thetable}{S\arabic{table}}
    \caption{
    Finite-size scaling results for the measurement-induced phase transition
    at all values of quasiperiodic potential $\lambda$.
    $\gamma_c^{(\mathrm{top})}$ and $\nu^{(\mathrm{top})}$ are extracted
    from the collapse of the topological entropy $S_\mathrm{top}$;
    $\gamma_c^{(L/2)}$ and $\nu^{(L/2)}$ from the half-system entanglement
    entropy $S_{L/2}$.
    System sizes used are $L=55,89,144,233,377$, with $\Delta=0.5$ and $J=1$.
    }
    \label{tab:S1}
    \begin{ruledtabular}
    \begin{tabular}{ccccc}
    $\lambda$ &
    $\gamma_c^{(\mathrm{top})}$ &
    $\nu^{(\mathrm{top})}$ &
    $\gamma_c^{(L/2)}$ &
    $\nu^{(L/2)}$ \\
    \hline
    $0$ &
      no transition &
      - &
      $1.458\pm 0.166$ &
      $2.356\pm 0.471$ \\
    $0.5$ &
      $1.826\pm 0.045$ &
      $2.485\pm 0.382$ &
      $2.273\pm 0.303$ &
      $2.552\pm 0.253$ \\
    $1$ &
      $2.010\pm 0.054$ &
      $1.719\pm 0.357$ &
      $2.357\pm 0.213$ &
      $2.430\pm 0.417$ \\
    $2$ &
      $1.814\pm 0.081$ &
      $2.452\pm 0.826$ &
      $2.083\pm 0.240$ &
      $2.210\pm 0.345$ \\
    $3$ &
      $1.225\pm 0.051$ &
      $2.101\pm 0.436$ &
      $1.569\pm 0.158$ &
      $2.912\pm 0.364$ \\
    $4$ &
      $0.686\pm 0.054$ &
      $1.740\pm 0.498$ &
      $1.102\pm 0.188$ &
      $3.202\pm 0.497$ \\
    $5$ &
      $0.164\pm 0.089$ &
      $1.664\pm 0.200$ &
      $0.443\pm 0.098$ &
      $3.145\pm 0.565$ \\
    $6$ &
      $0$ &
      - &
      $0$ &
      - \\
    $7$ &
      $0$ &
      - &
      $0$ &
      - \\
    \end{tabular}
    \end{ruledtabular}
\end{table*}

The systematic difference between $\gamma_c^{(\mathrm{top})}$ and $\gamma_c^{(L/2)}$ reflects the larger finite-size corrections in $S_{L/2}$ near the transition point. Specifically, $S_{L/2}$ systematically overestimates $\gamma_c$ compared to $S_\mathrm{top}$, which provides a cleaner crossing due to its sensitivity to the nonlocal correlations and smaller finite-size effects. We therefore use $\gamma_c^{(\mathrm{top})}$ as the primary estimate of the transition point throughout the paper.

For $\lambda=0$, $S_\mathrm{top}$ shows no crossing up to $L=377$, confirming the absence of a genuine MIPT in symmetry class~D NLSM~\cite{Fava2023}. We interpret the apparent crossing in $S_{L/2}$ at $\gamma_c^{(L/2)}=1.458\pm 0.166$ as a finite-size artifact that should disappear in the thermodynamic limit, as expected for a class-D system with no stable critical phase.

For $\lambda \geq 6$, both diagnostics give $\gamma_c=0$, indicating that strong quasiperiodic disorder completely suppresses the measurement-induced phase transition, consistent with the Zeno localization picture discussed in the main text. After extrapolating the transition boundary based on $S_{\rm top}$ collapses towards the $\gamma \to 0$ limit, we find that the critical phase is completely suppressed by the quasiperiodicity at around $\lambda \approx 5.3$.

For each diagnostic, the correlation length exponent $\nu$ is constant to within our numerical precision across all $\lambda$ values where a transition exists. Fitting our results to a constant ansatz yields $\nu^{(top)} = 1.86 \pm 0.14$ with reduced chi-squared of 0.934, and $\nu^{(L/2)} = 2.63 \pm 0.15$ with reduced chi-squared of 0.912 (excluding the $\lambda=0$ result). We associate the higher $\nu^{(L/2)}$ value with more pronounced finite-size effects in $S_{L/2}$.

\section{Effective non-Hermitian theory for the monitored Bogoliubov-de Gennes chain}
\label{sec:effective_theory}

The energy-density statistics discussed in the End Matter motivate a zero-energy reference description. We adopt this description for the orbital tails and derive its localization exponent. Our effective theory largely follows arguments from the number-conserving case~\cite{Singha2026}; however, important modifications and additional assumptions were required to establish the closed-form expressions and describe corrections to the localization length. We establish the correspondence with QSD orbitals conditionally on the tail assumptions stated in Sec.~\ref{sec:S2C.mult}. 


The dynamics of the conditioned pure state $|\Psi(t)\rangle$ under continuous measurement of local densities $n_j = c_j^\dagger c_j$ is described by the quantum state diffusion (QSD) equation:
\begin{align}
    d|\Psi\rangle &= \Big[-iH - \frac{\gamma}{2}\sum_j (n_j - \langle n_j\rangle)^2\Big]dt\,|\Psi\rangle
    + \sum_j \sqrt{\gamma}\,(n_j - \langle n_j\rangle)\,dW_j\,|\Psi\rangle,
    \label{eq:stochastic_eq2}
\end{align}
where $dW_j$ are independent Wiener increments with zero mean and variance $dt$, and $\gamma$ is the measurement strength. The quantum Zeno regime is defined by
\begin{equation}
    \gamma \gg J,\Delta,\qquad 
    \epsilon \equiv \frac{\sqrt{J^2+\Delta^2}}{\sqrt{\lambda^2+(\gamma/2)^2}} \ll 1.
    \label{Seq:zeno_hierarchy}
\end{equation}
In this limit, frequent measurements suppress coherent transport, and the state localizes around a manifold of pointer states. We choose an arbitrary reference pointer state, represented at the single-particle level by a dominant Nambu spinor $\Phi_0$ exponentially localized about site $j_0$. The instantaneous mode is expanded into a dominant localized spinor and fluctuations,
\begin{equation}
    \Psi_j(t) = \Phi_0(j) + \delta\Phi_j(t),\qquad 
    \delta\Phi_j(t) = \sum_{\alpha\ge1} c_\alpha(t)\,\Phi_\alpha(j),
    \label{eq:fluct_exp}
\end{equation}
with $\langle \Phi_0|\Phi_\alpha\rangle = 0$ and $c_\alpha = \mathcal{O}(\epsilon)$, and $\Psi_j(t)$ is the wave function evaluated at site $j$ at time $t$.

Our goal is to relate the QSD decay exponent to an explicit analytical expression. We first supply Secs.~\ref{sec:zeno_localization} and \ref{sec:S2B}, which discuss the reference profiles and core fluctuation estimates. Then the argument separates two steps: comparing QSD and reference tails under the assumptions of Sec.~\ref{sec:S2C}, which gives equality of the asymptotic exponents, and approximating the reference transfer exponent by a phase average in Sec.~\ref{sec:majorana_derivation}, finally yielding a closed form expression up to a controlled error $\mathcal{O}(\epsilon^2)$ for the localization length in Sec.~\ref{sec:closed_form}. Finally, Sec.~\ref{sec:localization_length_qsd} describes the numerical extraction from QSD orbitals.

\subsection{Localization of the dominant Nambu mode in the Zeno limit}
\label{sec:zeno_localization}

We study a reference profile whose tails satisfy $(H_{\mathrm{BdG}} - i \gamma \tau_z/2) \Phi_0 = 0$ outside the localization core. The core itself supplies the matching conditions for the decaying solutions. Comparison with QSD orbitals is made separately in Sec.~\ref{sec:S2C}.


\subsubsection{Effective non-Hermitian Hamiltonian in the Majorana basis}

The deterministic part of the QSD equation for $\Psi_j=(u_j,v_j)^T$ is
\begin{align}
    \dot{u}_j &= -i\Big[-J(u_{j+1}+u_{j-1}) + V_j u_j + \Delta(v_{j-1}-v_{j+1})\Big]
     -\frac{\gamma}{2}(1-2\langle n_j\rangle)u_j, \label{Seq:qsd_u_det}\\[4pt]
    \dot{v}_j &= -i\Big[J(v_{j+1}+v_{j-1}) - V_j v_j + \Delta(u_{j+1}-u_{j-1})\Big]
     +\frac{\gamma}{2}(1-2\langle n_j\rangle)v_j, \label{Seq:qsd_v_det}
\end{align}
where we omit the explicit time-dependence for clarity. Retaining the linear backbone of the drift and deferring the $\langle n_j\rangle$-dependence to the explicit nonlinear terms of Sec.~\ref{sec:nonlinear}, define $H_{\rm eff} = H_{\rm BdG} - i\tfrac{\gamma}{2}\sum_j c_j^\dagger c_j$, with
\begin{equation}
    H_{\rm BdG} = \sum_j\Big[V_j c_j^\dagger c_j - J(c_j^\dagger c_{j+1}+\text{h.c.}) + \Delta(c_jc_{j+1}+\text{h.c.})\Big].
\end{equation}
Passing to the Majorana basis $c_j=(a_j+ib_j)/2$, $c_j^\dagger=(a_j-ib_j)/2$, with $\{a_j,a_k\}=\{b_j,b_k\}=2\delta_{jk}$ and $\{a_j,b_k\}=0$, a direct substitution gives $c_j^\dagger c_j = \tfrac12 + \tfrac{i}{2}a_jb_j$ and, after collecting the hopping and pairing bilinears using the same anticommutation relations,
\begin{equation}
    H_{\rm eff} = \frac{i}{2}\sum_j\Big[(V_j-i\gamma/2)\,a_jb_j - (J+\Delta)\,a_{j+1}b_j + (J-\Delta)\,b_{j+1}a_j\Big],
    \label{Seq:Heff_majorana}
\end{equation}
up to an irrelevant additive constant. Eq.~\eqref{Seq:Heff_majorana} can be verified term by term from the fermionic Hamiltonian above and is exact.

\subsubsection{Zero-mode recursion}

A zero-energy Majorana operator $\Gamma = \sum_j(\psi_j a_j + \phi_j b_j)$ satisfies $[H_{\rm eff},\Gamma]=0$. Evaluating the commutator with the identity $[m_pm_q,m_r] = 2\delta_{qr}m_p - 2\delta_{pr}m_q$ (valid for any set of Majorana operators $\{m_i\}$ with $\{m_i,m_j\}=2\delta_{ij}$) and collecting the coefficients of $a_l$ and $b_l$ gives the coupled recursions
\begin{align}
    (J+\Delta)\psi_{j+1} + (J-\Delta)\psi_{j-1} - (V_j - i\gamma/2)\psi_j &= 0, \label{Seq:rec_psi}\\
    (J+\Delta)\phi_{j-1} + (J-\Delta)\phi_{j+1} - (V_j - i\gamma/2)\phi_j &= 0. \label{Seq:rec_phi}
\end{align}


Both recursions share identical structure; it suffices to analyze the scalar equation for $\psi_j$,
\begin{equation}
    (J+\Delta)\psi_{j+1} + (J-\Delta)\psi_{j-1} = (V_j - i\gamma/2)\psi_j.
    \label{Seq:scalar_recurrence}
\end{equation}
This is the same non-Hermitian tight-binding recursion encountered in the AAH problem, with effective hoppings $J\pm\Delta$ and imaginary on-site potential $-i\gamma/2$, which is what allows the localization analysis of the number-conserving case to transfer directly to the superconducting chain.

\subsubsection{Localization of the reference mode}
\label{sec:nonlinear}

We treat the localization core as a matching region and use Eq.~\eqref{Seq:scalar_recurrence} outside it. On the decaying right and left branches, respectively,
\begin{equation}
    \psi_j = \frac{J-\Delta}{V_j-i\gamma/2}\psi_{j-1} [1+\mathcal{O}(\epsilon^2)],
    \qquad
    \psi_j = \frac{J+\Delta}{V_j-i\gamma/2}\psi_{j+1} [1+\mathcal{O}(\epsilon^2)].
\end{equation}
Since $|V_j - i \gamma / 2| \ge \gamma / 2$, each outward step is suppressed by $\mathcal{O}(\epsilon)$. These relations describe the reference tails after matching to the core.
Repeating site by site yields the exponential bound
\begin{equation}
    |\psi_j^{(0)}| \le C_0\, e^{-|j-j_0|/\xi_0}, \qquad \xi_0^{-1} = \ln(1/\epsilon) + \mathcal{O}(1)
\end{equation}
for the slower reference tail, and the associated Nambu spinor inherits the same decay,
\begin{equation}
    |\Phi_0(j)| \le C_0'\, e^{-|j-j_0|/\xi_0}, \qquad |\Phi_0(j_0)| = \mathcal{O}(1).
    \label{eq:Phi0_decay}
\end{equation}
This characterizes the localization of the reference tails; the comparison with QSD orbitals is made in Sec.~\ref{sec:S2C}.

\bigskip
\subsubsection{Suppression of fluctuation modes at the core}
\label{sec:fluctuation}

We prove $|\Phi_\alpha(j_0)|=\mathcal{O}(\epsilon)$ for every mode $\Phi_\alpha$ orthogonal to $\Phi_0$, using the same perturbative construction as for the dominant mode: an unperturbed, site-localized seed corrected order by order in $\epsilon$.

Every stationary Nambu mode obeys the linear recursion [Eq.~\eqref{Seq:rec_psi}]
\begin{equation}
    (J+\Delta)\psi_{j+1} + (J-\Delta)\psi_{j-1} = (V_j - i\gamma/2)\,\psi_j,
    \label{Seq:rec_psi_repeat}
\end{equation}
which we solve for the amplitude at site $j$,
\begin{equation}
    \psi_j = \frac{(J+\Delta)\psi_{j+1} + (J-\Delta)\psi_{j-1}}{V_j - i\gamma/2}.
    \label{Seq:self_consistent}
\end{equation}
Consider a mode $\Phi_\alpha$ centered at a site $j_\alpha$, meaning $|\psi_{j_\alpha}|=\mathcal{O}(1)$. Write it as a seed plus correction,
\begin{equation}
    \psi_j^{(\alpha)} = \psi_j^{(\alpha,0)} + \delta\psi_j^{(\alpha)},
    \qquad
    \psi_j^{(\alpha,0)} \equiv \delta_{j,j_\alpha},
    \label{Seq:expansion}
\end{equation}
where the seed $\psi_j^{(\alpha,0)}$ is the fully localized, zeroth-order state (amplitude $1$ at $j_\alpha$, zero elsewhere), and $\delta\psi_j^{(\alpha)}$ collects everything generated once the hopping in Eq.~\eqref{Seq:self_consistent} spreads this seed onto neighboring sites.

Substituting Eq.~\eqref{Seq:expansion} into Eq.~\eqref{Seq:self_consistent} at a site $j\neq j_\alpha$ gives the exact relation
\begin{equation}
    \delta\psi_j^{(\alpha)} = \frac{(J+\Delta)\,\psi_{j+1}^{(\alpha)} + (J-\Delta)\,\psi_{j-1}^{(\alpha)}}{V_j-i\gamma/2}.
    \label{Seq:delta_eq}
\end{equation}
Because $|V_j-i\gamma/2|\sim\sqrt{\lambda^2+(\gamma/2)^2}$ is parametrically large compared with $J,\Delta$ [Eq.~\eqref{Seq:zeno_hierarchy}], the right-hand side of Eq.~\eqref{Seq:delta_eq} is smaller than whichever of $\psi_{j\pm1}^{(\alpha)}$ enters it by a factor
\begin{equation}
    \frac{J\pm\Delta}{V_j-i\gamma/2} = \mathcal{O}(\epsilon).
    \label{Seq:suppression_factor}
\end{equation}
This is the mechanism behind the result: Eq.~\eqref{Seq:delta_eq} expresses the amplitude at $j$ as an $\mathcal{O}(\epsilon)$ fraction of the amplitude at its neighbors, so amplitude cannot leak away from the seed without being suppressed by a factor of $\epsilon$ at every step.

We now build $\delta\psi_j^{(\alpha)}$ site by site, starting from the seed and moving outward, keeping at each step only the neighbor already determined at the previous (lower) order and dropping the not-yet-determined, farther neighbor -- consistent, since that neighbor is one step further from $j_\alpha$ and hence higher order by the same logic. For $j > j_\alpha$, Eq.~\eqref{Seq:delta_eq} with the $\psi_{j+1}^{(\alpha)}$ term dropped at this order gives
\begin{equation}
    \psi_j^{(\alpha)} \approx \frac{(J-\Delta)}{V_j - i\gamma/2}\,\psi_{j-1}^{(\alpha)},
    \qquad j > j_\alpha,
    \label{Seq:iterate_right}
\end{equation}
and for $j < j_\alpha$, dropping instead the $\psi_{j-1}^{(\alpha)}$ term,
\begin{equation}
    \psi_j^{(\alpha)} \approx \frac{(J+\Delta)}{V_j - i\gamma/2}\,\psi_{j+1}^{(\alpha)},
    \qquad j < j_\alpha.
    \label{Seq:iterate_left}
\end{equation}
Starting from $\psi_{j_\alpha}^{(\alpha)}=1$ and applying Eq.~\eqref{Seq:iterate_right} (or \eqref{Seq:iterate_left}) once per lattice step, each application multiplies the amplitude by a factor $\mathcal{O}(\epsilon)$ from Eq.~\eqref{Seq:suppression_factor}. After $n=|j-j_\alpha|$ steps,
\begin{equation}
    |\psi_j^{(\alpha)}| \le C\,\epsilon^{\,|j-j_\alpha|},
    \qquad
    |\Phi_\alpha(j)| \le C'\,\epsilon^{\,|j-j_\alpha|}.
    \label{Seq:generic_decay}
\end{equation}
This reproduces, for a general center $j_\alpha$, the same decay estimate used for $\Phi_0$ at $j_0$ -- the only difference is which site plays the role of the seed.

The dominant mode $\Phi_0$ is singled out among these solutions by the nonlinear stationarity condition of Sec.~\ref{sec:nonlinear}, which fixes its seed to site $j_0$. Any mode $\Phi_\alpha$ orthogonal to $\Phi_0$ is linearly independent of it and, being itself of the localized form \eqref{Seq:expansion}--\eqref{Seq:generic_decay}, has a seed at some site $j_\alpha$. Two solutions built from the same seed site would coincide up to normalization (the expansion \eqref{Seq:expansion}--\eqref{Seq:iterate_left} determines the mode uniquely once its center is fixed), so orthogonality forces $j_\alpha \neq j_0$, hence $|j_0-j_\alpha|\ge1$ on the lattice.

Evaluating Eq.~\eqref{Seq:generic_decay} for $\Phi_\alpha$ at the site $j_0$ occupied by the other mode, and using $|j_0-j_\alpha|\ge1$, gives
\begin{equation}
    |\Phi_\alpha(j_0)| \;\le\; C'\,\epsilon^{\,|j_0-j_\alpha|} \;\le\; C'\,\epsilon,
    \qquad \alpha \ge 1,
    \label{Seq:falpha_core}
\end{equation}
which is the claimed suppression. The bound is generically saturated by whichever fluctuation mode is seeded at a nearest-neighbor site of $j_0$ ($|j_0-j_\alpha|=1$); modes seeded further away are suppressed at $j_0$ by correspondingly higher powers of $\epsilon$. As an immediate corollary, the overlap $\langle\Phi_0|\Phi_\alpha\rangle$ -- built from the product of two such decaying tails meeting somewhere between $j_0$ and $j_\alpha$ -- is itself exponentially small, consistent with the (near-)orthogonality assumed for $\{\Phi_0,\Phi_\alpha\}$.

\bigskip
\noindent\textit{Remark.} The one-sided iterations \eqref{Seq:iterate_right}--\eqref{Seq:iterate_left} are self-consistent precisely because they always divide by the large diagonal entry $V_j-i\gamma/2$ rather than by the small hopping $J\pm\Delta$; dividing the other way would spuriously suggest growth away from the seed. As in the $\Phi_0$ case, this construction further assumes $V_{j_0}-i\gamma/2$ and $V_{j_\alpha}-i\gamma/2$ are not fine-tuned to near-coincidence, which holds generically for disordered or incommensurate $V_j$.

\subsection{Norm conservation and moment bound}
\label{sec:S2B}

Normalization bounds the total fluctuation weight. Combined with the reference profile's concentration at its dominant component, this gives a bound on the fluctuation there.

The quantum state diffusion evolution preserves the norm of the state
pathwise,
\begin{equation}
    \|\Psi(t)\|^2 = 1,
    \label{eq:normconservation}
\end{equation}
for every trajectory and at every time. This property is exact and does
not rely on the Zeno limit or on any linearization of the dynamics.

Using the orthonormal expansion
\begin{equation}
    \Psi_j = c_0\,\Phi_0(j) + \sum_{\alpha \geq 1} c_\alpha\,\Phi_\alpha(j),
    \label{eq:expansion}
\end{equation}
with
\begin{equation}
    \langle \Phi_\alpha | \Phi_\beta \rangle = \delta_{\alpha\beta},
    \qquad
    \langle \Phi_0 | \Phi_\alpha \rangle = 0,
    \label{eq:orthonorm}
\end{equation}
the norm-conservation condition~\eqref{eq:normconservation} becomes
\begin{equation}
    |c_0|^2 + \sum_{\alpha \geq 1} |c_\alpha|^2 = 1.
    \label{eq:normsum}
\end{equation}
Since every term is non-negative, each fluctuation amplitude is bounded
pathwise,
\begin{equation}
    |c_\alpha|^2 \leq 1
    \qquad (\alpha \geq 1).
    \label{eq:cbound}
\end{equation}
Taking the expectation over the stochastic trajectories gives the moment
bound
\begin{equation}
    \mathbb{E}|c_\alpha|^2 \leq 1 = \mathcal{O}(1)
    \qquad (\alpha \geq 1).
    \label{eq:momentbound}
\end{equation}
This bound is exact and independent of the linearized fluctuation
dynamics, the Ornstein-Uhlenbeck approximation, or the projected drift
matrix. It holds equally for particle-like and hole-like modes.

The Zeno localization of the fluctuation modes supplies the
complementary spatial bound. In the Zeno regime, the modes satisfy
\begin{equation}
    |\Phi_\alpha(j)| \leq C\,\epsilon^{|j - j_\alpha|},
    \qquad
    j_\alpha \neq j_0
    \qquad (\alpha \geq 1),
    \label{eq:modeloc}
\end{equation}
where $j_0$ is the center of the dominant mode.
Since orthogonality forces $j_\alpha \neq j_0$ for all $\alpha \geq 1$,
we have $|j_0 - j_\alpha| \geq 1$, so $|\Phi_\alpha(j_0)| \leq C\epsilon$
for all $\alpha \geq 1$.
Grouping modes by distance $n = |j_\alpha - j_0| \geq 1$ from the core,
and using $|c_\alpha| \leq 1$ from~\eqref{eq:cbound} and at most
$\kappa_0 = \mathcal{O}(1)$ modes per lattice site, the fluctuation
field at the dominant-mode center is bounded by
\begin{equation}
    |\delta\Psi(j_0)|
    \leq
    \sum_{\alpha \geq 1} |c_\alpha|\,|\Phi_\alpha(j_0)|
    \leq
    C \sum_{n=1}^{\infty} \kappa_0\,\epsilon^n
    =
    \frac{C \kappa_0\,\epsilon}{1 - \epsilon}
    =
    \mathcal{O}(\epsilon).
    \label{eq:corebound}
\end{equation}
The geometric series converges since $\epsilon \ll 1$.
On the localization core of $\Phi_0$, where the density
$a_j = |u_j^0|^2$ is non-negligible, this implies
\begin{equation}
    |\delta u_j| = \mathcal{O}(\epsilon)\,|u_j^0|.
    \label{eq:dubound}
\end{equation}
These bounds [the exact moment bound~\eqref{eq:momentbound} and the Zeno core bound~\eqref{eq:dubound}] control the potential estimate on the localization core. Comparing asymptotic tail exponents requires the separate assumptions introduced below.

\subsection{Effective QSD potential and effective Hamiltonian}
\label{sec:S2C}

We first estimate the reconstructed potential on the localization core. We then compare the asymptotic decay exponents using a multiplicative representation and explicit tail assumptions.

In the stationary reference approximation to the deterministic orbital equation, the particle component satisfies
\begin{equation}
    -J(u_{j+1}+u_{j-1})+V_ju_j+\Delta(v_{j-1}-v_{j+1})
    =
    i\frac{\gamma}{2}(1-2\langle n_j\rangle)\,u_j.
    \label{eq:stationarity}
\end{equation}
The scalar drift $-(\gamma/2) \sum_j \langle n_j \rangle^2 \, $ in Eq.~\ref{eq:stochastic_eq2} is absorbed into the overall normalization.

We define $V_j^{\rm QSD}(t)$ as the reconstructed particle-component potential from the bare BdG equation. Solving algebraically,
\begin{equation}
    V_j^{\rm QSD}(t)
    =
    \frac{J(u_{j+1}+u_{j-1})-\Delta(v_{j-1}-v_{j+1})}{u_j}.
    \label{eq:defV}
\end{equation}
This is exact and makes no assumption about the Zeno limit.

We expand $\Psi_j = c_0\Phi_0(j)+\sum_{\alpha\geq 1}
c_\alpha\Phi_\alpha(j)$ with $|c_0|^2+\sum_\alpha
|c_\alpha|^2=1$, and write $u_j=u_j^0+\delta u_j$,
$v_j=v_j^0+\delta v_j$ where
$\delta u_j=\sum_{\alpha\geq 1}c_\alpha u_j^\alpha$.

\subsubsection{Additive estimates on the localization core}
\label{sec:S2C.additive}

The reconstructed potential is a ratio of neighboring amplitudes to the local particle amplitude. At core sites where the denominator stays bounded away from zero, absolute fluctuation estimates therefore control the potential correction.

The dominant mode satisfies~\eqref{eq:stationarity} with $\langle n_j\rangle=a_j=|u_j^0|^2$. Rearranging gives the reference potential
\begin{equation}
    \frac{N_j^0}{u_j^0}
    \equiv
    \frac{J(u_{j+1}^0+u_{j-1}^0)
          -\Delta(v_{j-1}^0-v_{j+1}^0)}{u_j^0}
    =
    V_j-i\frac{\gamma}{2}+i\gamma a_j,
    \label{eq:N0overu0}
\end{equation}
which matches main-text Eq.~(7).

Write $u_j=c_0u_{0j}+\delta u_j$ and $v_j=c_0v_{0j}+\delta v_j$, consistently with the normalized decomposition. Define
\begin{equation}
    \delta N_j = J(\delta u_{j+1}+\delta u_{j-1})
    -\Delta(\delta v_{j-1}-\delta v_{j+1}).
\end{equation}

Linearity of the numerator in Eq.~\eqref{eq:defV} gives the exact identity
\begin{equation}
    V_j^{\mathrm{QSD}}
    = V_{0j} + \frac{\delta N_j-V_{0j}\delta u_j}{c_0u_{0j}+\delta u_j}, \qquad V_{0j} \equiv \frac{N_j^0}{u_j^0}.
\end{equation}
At core sites where the actual particle amplitude is bounded away from zero, locality gives
\begin{equation}
    |\delta N_j| \le \sqrt{2(J^2+\Delta^2)}\,\|\delta\Phi\|.
\end{equation}
Together with the core fluctuation estimate, $\|\delta\Phi\|\le1$, and the stated Zeno scaling, this yields
\begin{equation}
    |\delta V_j^{\rm stoch}|
    =\mathcal{O}(\sqrt{J^2+\Delta^2}).
    \label{eq:dVbound}
\end{equation}
This estimate applies to the core and does not establish a uniform bound along the tails.
Since $|V_j-i\gamma/2|=\mathcal{O}(\epsilon^{-1}
\sqrt{J^2+\Delta^2})$, the relative correction is
\begin{equation}
    \frac{|\delta V_j^{\rm stoch}|}{|V_j-i\gamma/2|}
    =\mathcal{O}(\epsilon),
    \label{eq:relcorrection}
\end{equation}
so the stochastic correction is subleading. The effective
potential therefore takes the form
\begin{equation}
    V_j^{\rm QSD}(t)
    =
    V_j-i\frac{\gamma}{2}
    +i\gamma a_j
    +\delta V_j^{\rm stoch}(t),
    \qquad
    |\delta V_j^{\rm stoch}|
    =\mathcal{O}(\sqrt{J^2+\Delta^2}).
    \label{eq:Vfinal}
\end{equation}
The additive estimates above apply on the localization core. The core contribution $i\gamma a_j$ decays away from the localization center. The asymptotic decay exponent is analyzed separately below using the multiplicative representation.

\subsubsection{Multiplicative representation on the localized tails}
\label{sec:S2C.mult}

In the tails, a small absolute fluctuation can still be comparable to the exponentially small reference amplitude. We therefore describe the deformation through the relative factor $\Psi_a / \Phi_a = 1 + \eta_a$ (we do not require $\eta_a$ itself to be small).

Let $a=(j,\sigma)$ with $\sigma\in\{u,v\}$ and define the relative fluctuation
\begin{equation}
    \Psi_a = \Phi_a(1+\eta_a),
    \qquad
    \eta_a = \frac{\delta\Psi_a}{\Phi_a}.
    \label{eq:multPhi}
\end{equation}
Here $\Phi_a$ is the reference localized profile whose tails satisfy the equation $(H_{\text{BdG}} - i \gamma \tau_z / 2) \Phi = 0$ away from the localization core. We impose the following conditions along each tail, each physically motivated and illustrated by the numerical diagnostics below.

\begin{figure}[t]
\centering
\includegraphics[width=\textwidth]{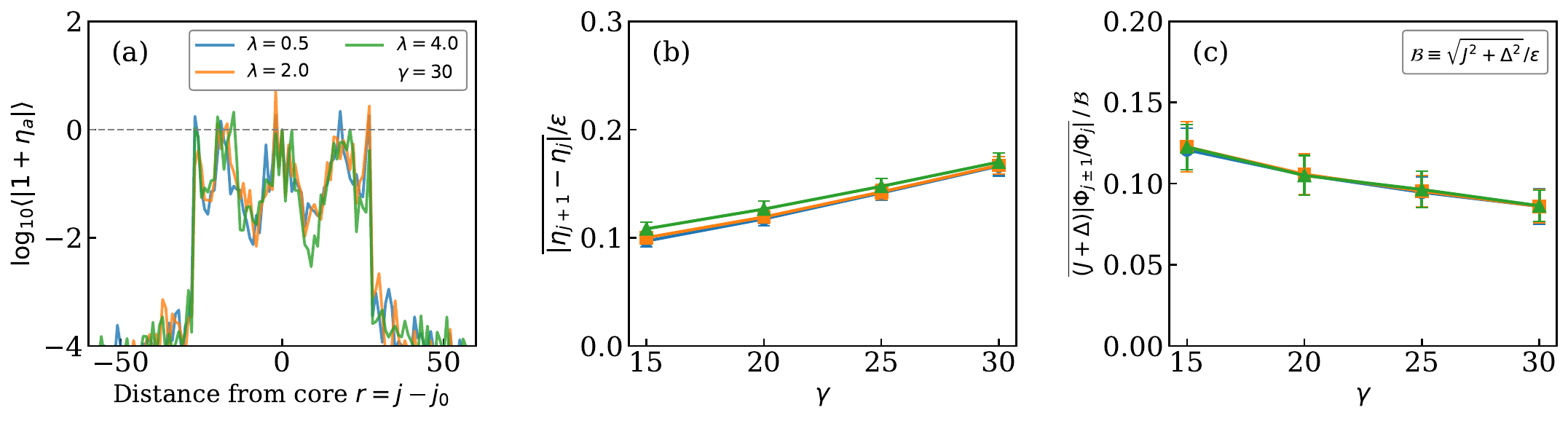}
\caption{
Numerical diagnostics for the reference profiles selected under assumptions A1--A3, for $L=233$, $\Delta=0.5$, $J=1.0$, averaged over 128 disorder realizations per $(\lambda,\gamma)$.
\textbf{(a)}~A1: the relative amplitude $\log_{10}\langle|1+\eta_a|\rangle$ plotted against the distance from the localization core $r = j - j_0$ at $\gamma=30$ for various $\lambda$. Over the resolved tail region, the plotted mean relative amplitude is consistent with the assumed behavior. The bounded oscillations reflect physical wavefunction nodes and quasiperiodic interference. Far away from the localization core, the QSD profiles reach numerical precision, and thus $\log_{10}\langle|1+\eta_a|\rangle$ sharply changes near $|r|\approx 25$; results for $|r|\gtrsim 25$ are therefore numerical noise.
\textbf{(b)}~A2: the mean increment ratio $\overline{|\eta_{j+1} - \eta_j|} / \epsilon$ remains of order unity over the sampled parameter range, consistent with the assumed $\mathcal{O}(\epsilon)$ increment scaling.
\textbf{(c)}~A3: the mean amplitude ratio $\overline{(J + \Delta) |\Phi_{j\pm 1} / \Phi_j|} / \mathcal{B}$, where $\mathcal{B} \equiv \sqrt{J^2+\Delta^2} / \epsilon$, is well below unity for all $(\lambda,\gamma)$. The plotted mean ratio is consistent with the scale assumed in A3 over the sampled range.
}
\label{fig:assumptions}
\end{figure}

\medskip
\noindent\textit{A1 (uniform nonvanishing relative amplitude):}
The reference profiles used in Eq.~\eqref{eq:multPhi} are nonzero, and $|1+\eta_a| \ge c_*> 0$, where $c_*$ is independent of distance, system size, and $\epsilon$.
Numerically, the reference profiles are approximated by fitting the tails to an exponentially decaying form, and therefore the assumption is expected to hold [see Fig.~\ref{fig:assumptions}(a); note that far away from the core, the QSD wave functions reach numerical precision.

\medskip
\noindent\textit{A2 (uniformly bounded increments):}
Whenever $h_{ab}\ne0$, $|\eta_a-\eta_b|\le C\epsilon$, where $C$ is independent of distance, system size, and $\epsilon$.
Numerically, $\overline{|\eta_{j+1}-\eta_j|}/\epsilon$ is approximately constant across $\gamma\in\{15,20,25,30\}$ and $\lambda\in\{0.5,2.0,4.0\}$, consistent with $\mathcal{O}(\epsilon)$ scaling (Fig.~\ref{fig:assumptions}).

\medskip
\noindent\textit{A3 (no resonances):}
$\sum_{b\neq a}|h_{ab}||\Phi_b/\Phi_a|
<c\sqrt{J^2+\Delta^2}/\epsilon$ for all $a$,
with $c=\mathcal{O}(1)$.
At core sites $|\Phi_b/\Phi_a|=\mathcal{O}(\epsilon)$;
at tail sites the ratio toward the core is at most
$\mathcal{O}(\epsilon^{-1})$; both follow from
Sec.~\ref{sec:zeno_localization}. Numerically, the mean ratio
is approximately constant across all parameters and well below the bound (Fig.~\ref{fig:assumptions}).

\medskip

A1 controls the denominator, A2 controls differences between neighboring relative fluctuations, and A3 controls the weighted reference-amplitude ratios.

Substituting into $\sum_b h_{ab}\Psi_b
=V_a^{\rm QSD}\Psi_a$ and writing
$\eta_b=\eta_a+(\eta_b-\eta_a)$,
\begin{equation}
    V_a^{\rm QSD}
    =
    \frac{(h_{\rm BdG}\Phi)_a}{\Phi_a}
    +
    \underbrace{
    \frac{
        \sum_{b\neq a}h_{ab}\Phi_b(\eta_b-\eta_a)
    }{
        \Phi_a(1+\eta_a)
    }}_{\delta V_a^{\rm stoch}}.
\end{equation}
Applying A1--A3, we obtain
\begin{equation}
    |\delta V_a^{\rm stoch}|
    \leq
    \frac{C\epsilon\cdot c\sqrt{J^2+\Delta^2}/\epsilon}
         {|1+\eta_a|}
    =\mathcal{O}(\sqrt{J^2+\Delta^2}).
    \label{eq:multbound}
\end{equation}

Note that the additive and multiplicative identities are algebraically equivalent when written for the same reference $\Phi = c_0 \Phi_0$, with $\delta \Psi_a = \Phi_a \eta_a$. This equivalence does not extend the additive core estimate to the tails: the potential estimate uses A1--A3, whereas the decay-exponent comparison uses A1--A2.

\subsubsection{Effective Hamiltonian and summary}
\label{sec:S2C.Heff}

The reference operator governing the tails is
\begin{equation}
    H_{\mathrm{eff}}\equiv H_{\mathrm{BdG}}-i\frac{\gamma}{2}\tau_z.
    \label{eq:Heff}
\end{equation}
The localized core contribution determines the matching to the core and is omitted from this tail operator. The additive calculation establishes a core estimate, while the multiplicative potential estimate is conditional on A1--A3.

We now compare decay exponents. Along either tail, with $r$ the distance from a fixed starting site, A1 and A2 give
\begin{equation}
    0 < c_\ast \le |1 + \eta_{j_0\pm r,\sigma}|
    \le M+C\epsilon r,
\end{equation}
where $M = \max_\sigma |1 + \eta_{j_0,\sigma}|$. Consequently, for the site-resolved Nambu amplitudes,
\begin{equation}
    0 < c_*\|\Phi_{j_0\pm r}\|
    \le \|\Psi_{j_0\pm r}\|
    \le (M+C\epsilon r)\|\Phi_{j_0\pm r}\|.
\end{equation}
Their asymptotic decay exponents therefore coincide:
\begin{equation}
    \kappa_{\mathrm{QSD}}-\kappa_{\mathrm{eff}} = -\lim_{r\to\infty}\frac{1}{r}     \ln\frac{\|\Psi_{j_0\pm r}\|}{\|\Phi_{j_0\pm r}\|} = 0.
\end{equation}
Here $\kappa_{\mathrm{eff}}$ denotes the reference tail-decay exponent, using the transfer-branch identification in Sec.~\ref{sec:majorana_derivation}. The relative amplitude stays bounded away from zero and grows at most linearly with distance. Its logarithm therefore grows sublinearly and does not contribute to the asymptotic decay exponent. Combining this identity with the transfer-matrix expansion in Sec.~\ref{sec:closed_form} gives
\begin{equation}
    \kappa_{\mathrm{QSD}}
    =\kappa_{\mathrm{eff}}
    =\kappa_I+\mathcal{O}(\epsilon^2).
\end{equation}
For positive $\kappa_I$,
\begin{equation}
    \xi_{\mathrm{QSD}}
    =\frac1{\kappa_I}
    +\mathcal{O}\Big(\frac{\epsilon^2}{\kappa_I^2}\Big).
\end{equation}
Thus, under the stated tail assumptions, the error relative to the phase-averaged expression comes from the transfer-matrix approximation and is $\mathcal{O}(\epsilon^2)$.

\subsection{Majorana representation and transfer matrix derivation}
\label{sec:majorana_derivation}

This subsection supplies the algebra underlying the reference recursions used in Sec.~\ref{sec:zeno_localization}. We derive the transfer matrix and identify which transfer exponent controls the slower decaying Nambu tail. 

We begin with the one-dimensional fermionic Hamiltonian defined in
the main text [Eq.~\eqref{eq:H}]. To express this Hamiltonian in terms of
Majorana fermions, we introduce the operators
\begin{equation}
    c_j = \frac{a_j + i b_j}{2}, \qquad
    c_j^\dagger = \frac{a_j - i b_j}{2},
\end{equation}
which satisfy the Majorana algebra
\begin{equation}
    \{a_j,a_k\} = 2\delta_{jk}, \quad
    \{b_j,b_k\} = 2\delta_{jk}, \quad
    \{a_j,b_k\} = 0.
\end{equation}

Substituting these expressions into the Hamiltonian and simplifying, one finds that the onsite term reduces, up to an irrelevant constant shift, to
\begin{equation}
    H_{\mathrm{onsite}} = \frac{i}{2} \sum_j V_j a_j b_j,
\end{equation}
while the hopping and pairing contributions become
\begin{align}
    H_{\mathrm{hop}} &= -\frac{iJ}{2}\sum_j
    \left(a_{j+1}b_j + a_j b_{j+1}\right), \\
    H_{\mathrm{pair}} &= -\frac{i\Delta}{2}\sum_j
    \left(a_{j+1}b_j - a_j b_{j+1}\right).
\end{align}
Combining these terms, the Hamiltonian takes the compact form
\begin{equation}
    H = \frac{i}{2}\sum_j
    \Big[
    V_j a_j b_j
    - (J+\Delta)a_{j+1}b_j
    + (J-\Delta)b_{j+1}a_j
    \Big].
    \label{eq:H_majorana_clean}
\end{equation}

We now include the measurement-induced non-Hermitian contribution
\begin{equation}
    H_{\mathrm{eff}} = H - i\frac{\gamma}{2}\sum_j n_j,
\end{equation}
where the density operator is
\begin{equation}
    n_j = \frac{1}{2} + \frac{i}{2} a_j b_j.
\end{equation}
Discarding the constant part, this term simply renormalizes the onsite potential as
\begin{equation}
    V_j \;\longrightarrow\; V_j - i\frac{\gamma}{2}.
\end{equation}

The effective Hamiltonian therefore becomes
\begin{equation}
    H_{\mathrm{eff}} =
    \frac{i}{2} \sum_j
    \Big[
    (V_j - i\gamma/2) a_j b_j
    - (J+\Delta) a_{j+1} b_j
    +(J-\Delta) b_{j+1} a_j
    \Big].
    \label{eq:Heff_majorana_clean}
\end{equation}
To identify zero-energy modes, we consider a linear fermionic operator expanded in the Majorana basis
\begin{equation}
    \Gamma = \sum_j \left(\psi_j a_j + \phi_j b_j\right),
\end{equation}
and impose the condition
\begin{equation}
    [H_{\mathrm{eff}}, \Gamma] \approx 0.
\end{equation}

Using the Majorana commutation relations, a direct calculation from
$\{a_j,a_k\}=2\delta_{jk}$, $\{b_j,b_k\}=2\delta_{jk}$, $\{a_j,b_k\}=0$ gives
\begin{equation}
    [i a_m b_n, a_k] = -2i\,\delta_{mk} b_n, \qquad
    [i a_m b_n, b_k] = 2i\,\delta_{nk} a_m.
    \label{eq:majorana_comm}
\end{equation}
(Explicitly, for the second relation: $[a_mb_n,b_n] = a_mb_n b_n - b_n a_m b_n = a_m - (-a_m) = 2a_m$, using $b_n^2=1$ and $b_na_m=-a_mb_n$; hence $[ia_mb_n,b_k]=+2i\delta_{nk}a_m$, with the same overall sign as the $\gamma_i\gamma_j$-type identity $[\gamma_i\gamma_j,\gamma_k]=2\delta_{jk}\gamma_i-2\delta_{ik}\gamma_j$ for anticommuting Majorana operators.)

Using \eqref{eq:majorana_comm}, the commutator can be evaluated explicitly. After collecting terms, it takes the form
\begin{equation}
    [H_{\mathrm{eff}}, \Gamma]
    =  \sum_j \left( A_j\, b_j + B_j\, a_j \right).
\end{equation}
Since the Majorana operators $\{a_j\}$ and $\{b_j\}$ form an independent operator basis, the zero-mode condition requires
\begin{equation}
    A_j = 0, \qquad B_j = 0 \quad \forall j.
\end{equation}
This leads to the coupled recursion relations
\begin{align}
    \label{eq:reccurence_eqsn}
    (J+\Delta)\psi_{j+1}
    + (J-\Delta)\psi_{j-1}
    - (V_j - i\gamma/2)\psi_j &= 0, \\
    (J+\Delta)\phi_{j-1}
    + (J-\Delta)\phi_{j+1}
    - (V_j - i\gamma/2)\phi_j &= 0.
\end{align}
These equations describe right-moving and left-moving modes, respectively.
Although they appear distinct, they in fact lead to identical localization properties.

To see this, consider a spatial inversion $j \to -j$ and define $\tilde{\phi}_j = \phi_{-j}$. The backward equation then transforms into
\begin{equation}
    (J+\Delta)\tilde{\phi}_{j+1}
    + (J-\Delta)\tilde{\phi}_{j-1}
    - (V_{-j} - i\gamma/2)\tilde{\phi}_j = 0,
\end{equation}
which has the same structure as the forward equation up to the replacement $V_j \to V_{-j}$.

For quasiperiodic potentials of the form $V_j = \lambda \cos(2\pi \alpha j + \phi)$, this corresponds to a phase shift $\phi \to -\phi$ (with disorder average), which does not affect the Lyapunov exponent due to ergodicity. 
Therefore,
\begin{equation}
    \lambda_{\mathrm{forward}} = \lambda_{\mathrm{backward}}.
\end{equation}
It is thus sufficient to analyze only the forward equation. Rewriting it in a recursive form,
\begin{equation}
    \psi_{j+1} = a_j \psi_j + b\,\psi_{j-1},\qquad
    a_j = \frac{V_j - i\gamma/2}{J+\Delta}, \qquad
    b = -\frac{J-\Delta}{J+\Delta}.
    \label{eq:recursion_psi}
\end{equation}
Alternatively, this can be cast into a transfer-matrix form,
\begin{equation}
    \begin{pmatrix}
    \psi_{j+1} \\
    \psi_{j}
    \end{pmatrix}
    =
    T_j
    \begin{pmatrix}
    \psi_j \\
    \psi_{j-1}
    \end{pmatrix},
\end{equation}
with transfer matrix
\begin{equation}
    T_j =
    \begin{pmatrix}
    a_j
    & b \\
    1 & 0
    \end{pmatrix}.
\end{equation}

The localization properties are then characterized by the Lyapunov exponent
\begin{equation}
    \kappa_T = \lim_{L \to \infty} \frac{1}{L}
    \log \Big\| \prod_{j=1}^{L} T_j \Big\|.
    \label{K_Teqsn}
\end{equation}
This establishes the transfer matrix description. Since $\det T_j = (J-\Delta)/(J+\Delta)$ for $0 \le \Delta < J$, the two transfer exponents are $\kappa_T$ and $\ln[(J-\Delta) / (J+\Delta)] - \kappa_T$. The directional decay rates are therefore $\kappa_T$ and $\kappa_T - \ln[(J-\Delta) / (J+\Delta)]$, interchanged between the two Majorana channels. The Nambu tail has $\kappa_{\text{eff}} \approx \kappa_T$ when its matching to the core gives a nonzero coefficient of the slower branch.

\subsection{Closed-form localization length in the quantum Zeno regime}
\label{sec:closed_form}

In this section, we derive a closed-form expression for the Lyapunov exponent and the associated localization length for a non-Hermitian quasiperiodic system in the quantum Zeno regime. We first approximate the transfer exponent by the spatial average of the local logarithmic growth, and then evaluate that average for the quasiperiodic potential.

As per Eq.~\eqref{K_Teqsn}, the Lyapunov exponent is defined from the transfer-matrix product,
\begin{equation}
    \kappa_T = \lim_{n\to\infty} \frac{1}{n}\ln \|T_n \cdots T_1\|.
\end{equation}
In general, this quantity cannot be reduced to a simple phase average due to directional mixing in the matrix product.

We take large $\gamma$ limit at fixed $J,\Delta$, with $\lambda/\gamma$ bounded. Then the coefficients are uniformly large:
\begin{equation}
    |a_j| = \left|\frac{V_j-i\gamma/2}{J+\Delta}\right| \ge\frac{\gamma}{2(J+\Delta)},
    \qquad b=-\frac{J-\Delta}{J+\Delta} = \mathcal{O}(1).
\end{equation}
On the expanding solution used to compute the transfer-matrix Lyapunov exponent, define $R_j=\psi_j/\psi_{j-1}$. The exact recursion is
\begin{equation}
    R_{j+1} = a_j + \frac{b}{R_j}.
\end{equation}
On the expanding branch, $R_j \sim a_{j-1} = O(\epsilon^{-1})$. The feedback term $b / R_j$ is therefore $\mathcal{O}(\epsilon)$, giving a relative correction $\mathcal{O}(\epsilon^2)$ to the leading term $a_j$. Iterating in the uniformly small inverse coefficients gives
\begin{equation}
    R_{j+1} = a_j\left[1+\frac{b}{a_j a_{j-1}}+\mathcal{O}(\epsilon^4) \right].
\end{equation}
The Lyapunov exponent is therefore
\begin{equation}
    \kappa_T = \langle \ln|R_{j+1}| \rangle,
\end{equation}
and hence
\begin{equation}
    \kappa_T = \kappa_I +     \text{Re} \left\langle\frac{b}{a_j a_{j-1}} \right\rangle + \mathcal{O}(\epsilon^4) = \kappa_I + \mathcal{O}(\epsilon^2),
    \qquad \kappa_I = \langle \ln|a_j| \rangle.
\end{equation}





\begin{figure}[tb]
    \centering
    \includegraphics[width=0.5\columnwidth]{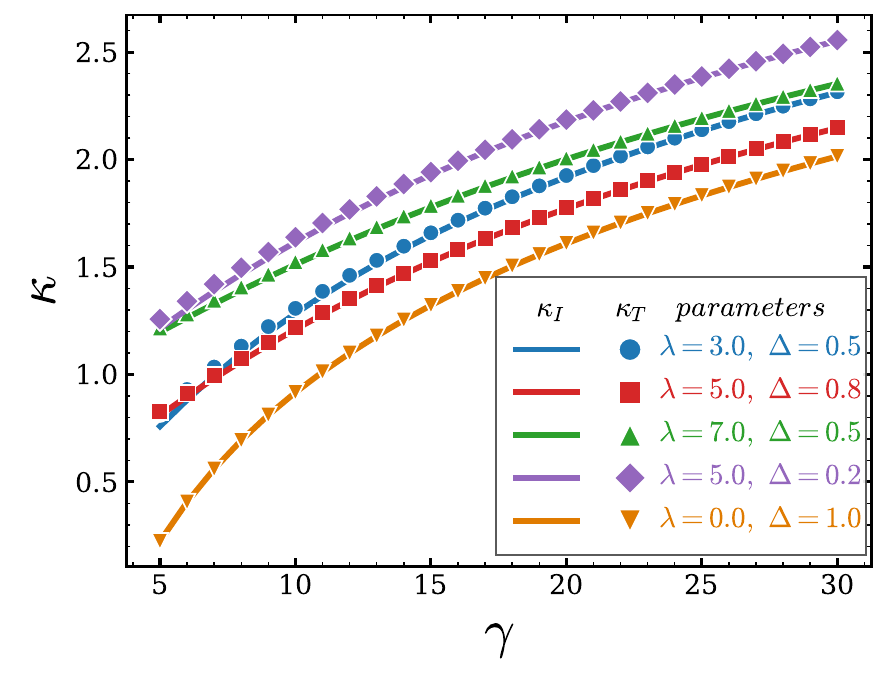}
    \caption{Lyapunov exponent $\kappa$ vs.\ measurement strength $\gamma$,
    comparing the analytical phase-averaged estimate $\kappa_I$ [solid lines,
    Eq.~\ref{K_Ieqsn}] against the numerically extracted transfer-matrix Lyapunov
    exponent $\kappa_T$ [symbols, Eq.~\ref{K_Teqsn}], for several parameter sets
    $(\lambda,\Delta)$ as indicated in the legend. The quantitative agreement
    across the full range of $\gamma$ confirms that directional mixing in the
    transfer-matrix product is perturbatively suppressed in the quantum Zeno
    regime, so that $\kappa_T \to \kappa_I$ to $\mathcal{O}(\epsilon^2)$.}
    \label{fig:kappaI_kappaT}
\end{figure}
We now evaluate the phase-averaged quantity $\kappa_I$. Using the explicit form of $a_j$, we obtain
\begin{equation}
    \kappa_I = 
    \left\langle 
    \ln\left|\frac{V_j - i\gamma/2}{J+\Delta}\right|
    \right\rangle
    =
    \left\langle \ln|V_j - i\gamma/2| \right\rangle - \ln(J+\Delta).
\end{equation}

For a quasiperiodic potential of the form
\begin{equation}
    V_j = \lambda \cos(2\pi\alpha j + \phi),
\end{equation}
with irrational $\alpha$, the sequence $\theta_j = 2\pi\alpha j + \phi$ densely fills the interval $[0,2\pi)$. As a result, spatial averages over the lattice can be replaced by averages over the phase $\theta$, which allows us to write
\begin{equation}
    \left\langle f(V_j)\right\rangle 
    = \frac{1}{2\pi}\int_0^{2\pi} f(\lambda\cos\theta)\, d\theta.
\end{equation}
Applying this to the logarithm, we obtain
\begin{equation}
    \kappa_I =
    \frac{1}{2\pi}\int_0^{2\pi} 
    \ln\left|\lambda\cos\theta - i\frac{\gamma}{2}\right| d\theta
    - \ln(J+\Delta).
\end{equation}

This integral can be evaluated analytically using standard results for logarithmic averages, yielding the closed-form expression
\begin{equation}
    \kappa_T =
    \ln\left(
    \frac{\sqrt{\lambda^2 + (\gamma/2)^2} + \gamma/2}
    {2(J+\Delta)}
    \right) + \mathcal{O}(\epsilon^2).
    \label{K_Ieqsn}
\end{equation}

The localization length is defined as the inverse Lyapunov exponent,
\begin{equation}
    \xi = \frac{1}{|\kappa_T|},
\end{equation}
and therefore, in the Zeno regime, it is given by
\begin{equation}
    \xi = \frac{1}{|\kappa_I|} + \mathcal{O}(\epsilon^2).
\end{equation}

As shown in Fig.~\ref{fig:kappaI_kappaT}, the analytical expression $\kappa_I$ agrees with the numerically computed $\kappa_T$ within numerical precision in the quantum Zeno regime. This confirms that the suppression of directional mixing renders the transfer-matrix product effectively multiplicative, leading to an asymptotically exact phase-averaged description of localization.

\subsection{Localization length from QSD}
\label{sec:localization_length_qsd}

For each unscrambled Bogoliubov orbital $n$, where unscrambling is done using Ref.~\cite{Szyniszewski2024}, we compute the site-resolved quasiparticle density
\begin{equation}
    \rho_n(j) = |U_{jn}|^2 + |V_{jn}|^2,
\end{equation}
and identify its localization center $j_0 = \arg\max_j \rho_n(j)$.

To extract the exponential tail of each orbital, we re-center $\rho_n$ at $j_0$ and separate it into a left and a right branch, measured in terms of the signed distance $y$ from the localization center ($y>0$ to the right, $y<0$ to the left). Along each branch we discard sites with vanishing weight and take the square-root amplitude $\sqrt{\rho_n(j)}$, retaining only its logarithm,
\begin{equation}
    \ell_n(y) = \ln \sqrt{\rho_n(j_0 + y)},
\end{equation}
truncated below a fixed cutoff $\ell_{\min}$ to remove numerical noise floor contributions far in the tail. Orbitals for which fewer than a minimum number of points survive this cut are discarded. This is repeated independently for the left and right branches of every orbital in every trajectory.

For a fixed set of parameters $(L,\lambda,\Delta,\gamma)$, the shifted log-amplitude tails $\ell_n(y)$ are collected over all retained orbitals and over $N$ independent QSD trajectories (seeds). At each distance $y$ we compute the sample mean and the standard error of the mean (SEM) across the pooled set of tails,
\begin{equation}
    \overline{\ell}(y) = \big\langle \ell_n(y) \big\rangle, \qquad
    \sigma_{\overline{\ell}}(y) = \frac{\mathrm{std}\big[\ell_n(y)\big]}{\sqrt{N_{\mathrm{eff}}(y)}},
\end{equation}
where $N_{\mathrm{eff}}(y)$ is the number of tails contributing a value at distance $y$ (tails of unequal length are aligned at $y=0$ and padded with missing values beyond their extent). This produces a single averaged log-amplitude profile $\overline{\ell}(y) \pm \sigma_{\overline{\ell}}(y)$ for each direction (left/right).

Within a chosen fitting window $[y_{\min},y_{\max}]$ (excluding points below the noise cutoff), the averaged tail is well described by a linear decay in $y$,
\begin{equation}
    \overline{\ell}(y) \approx a - \frac{y}{\xi},
\end{equation}
motivated by the expected exponential localization $\rho_n(j)\sim e^{-2|j-j_0|/\xi}$. We fit $a$ and $\xi^{-1}$ by weighted least squares, using $\sigma_{\overline{\ell}}(y)$ as the fit weights, to obtain the localization length $\xi$ for each direction. To assess the sensitivity of the extracted $\xi$ to the choice of fitting window, we repeat the fit after progressively excluding the points closest to $y=0$ (a window scan), and take the standard deviation of the resulting $\xi$ values across window choices as an additional \emph{window} uncertainty, combined in quadrature with the statistical fit uncertainty to give the total error $\delta\xi$.

Left- and right-branch estimates $\xi_L \pm \delta\xi_L$ and $\xi_R \pm \delta\xi_R$ are then combined into a single localization length via an inverse-variance weighted average,
\begin{equation}
    \xi_{\mathrm{avg}} = \frac{\xi_L/\delta\xi_L^2 + \xi_R/\delta\xi_R^2}{1/\delta\xi_L^2 + 1/\delta\xi_R^2}, \qquad
    \delta\xi_{\mathrm{avg}} = \left(\frac{1}{\delta\xi_L^2} + \frac{1}{\delta\xi_R^2}\right)^{-1/2}.
\end{equation}
This procedure is repeated independently for each value of $\gamma$ (and $\lambda$), yielding the localization length as a function of the measurement rate reported in the main text.